\pdfoutput=1
\documentclass[twocolumn]{autart}

\usepackage[sort&compress]{natbib}
\usepackage{amsmath,amssymb}
\usepackage{mathtools}
\usepackage{graphicx}
\usepackage{xcolor}
\usepackage{subfig}
\usepackage{tikz}
\usetikzlibrary{arrows.meta,positioning}
\usepackage{pgfplots}
\usepackage[font=scriptsize]{caption}
\usepackage{url}

\newtheorem{theorem}{Theorem}

\newtheorem{proposition}{Proposition}
\newtheorem{assumption}{Assumption}

\newtheorem{problem}{Problem}
\newtheorem{corollary}{Corollary}
\newtheorem{example}{Example}

\AtBeginDocument{%
  \setlength{\abovedisplayskip}{4pt}%
  \setlength{\belowdisplayskip}{4pt}%
  \setlength{\abovedisplayshortskip}{3pt}%
  \setlength{\belowdisplayshortskip}{3pt}%
}

\usepackage{tikz}
\usetikzlibrary{arrows.meta,positioning}
\usepackage{pgfplots}

\newcommand{\ba}{\begin{array}}
\newcommand{\ea}{\end{array}}

\newcommand{\be}{\begin{equation}}
\newcommand{\ee}{\end{equation}}
\newcommand{\beu}{\begin{equation*}}
\newcommand{\eeu}{\end{equation*}}

\newcommand{\ds}{\displaystyle}

\newcommand{\mc}{\mathcal}

\newcommand{\R}{\mathbb{R}}

\newcommand{\diag}{{\rm diag}\,}

\newcommand{\V}{\mathcal{V}}
\newcommand{\G}{\mathcal{G}}

\newcommand{\X}{\mathcal{Y}}
\newcommand{\Kset}{\mathcal{K}}
\renewcommand{\!}{\mspace{-2mu}}

\graphicspath{{figs}}

\begin{document}
    \begin{frontmatter}
    \title{Competition, Switching, and Control in Coupled Adoption-Opinion Dynamics}
    \thanks[footnoteinfo]{Corresponding author: Martina Alutto. \\
    This work was supported in part by Digital~Futures,~Stockholm,~Sweden, the Wallenberg AI, Autonomous Systems and Software Program (WASP) funded by the Knut and Alice Wallenberg Foundation, and by the European Union -- Next Generation EU, Mission 4, Component 1, under the PRIN project \emph{TECHIE: A control and network-based approach for fostering the adoption of new technologies in the ecological transition}, Cod. 2022KPHA24, CUP Master: D53D23001320006, CUP: B53D23002760006.}
    
    \author[KTH]{Martina Alutto\thanksref{footnoteinfo}}\ead{alutto@kth.se},    
    \author[CNR]{Fabrizio Dabbene}\ead{fabriziodabbene@cnr.it},               
    \author[KTH]{Angela Fontan}\ead{angfon@kth.se},
    \author[KTH]{Karl H. Johansson}\ead{kallej@kth.se},
    \author[CNR]{Chiara Ravazzi}\ead{chiararavazzi@cnr.it}
    
    \address[KTH]{Department of Decision and Control Systems, School of Electrical Engineering and Computer Science, KTH Royal Institute of Technology, Stockholm, Sweden.}                                           
    \address[CNR]{Institute of Electronics, Computer and Telecommunication Engineering, National Research Council of Italy, Politecnico di Torino, 10129 Torino, Italy.}     
    
    
    \begin{abstract}   
    We propose a Multi-technology AdopTion-opinion co-Evolution (MATE) model to describe the diffusion of $m \geq 2$ competing alternatives within a networked population whose opinions evolve under social influence and the feedback generated by adoption. Communities interact through two layers: a physical network governing exposure to adoption, and a social network shaping opinion evolution through social imitation. Individuals can adopt a technology, eventually becoming dissatisfied at a rate reflecting the quality of the technology, and switch to an alternative at a rate driven by their opinion of that alternative.
    For an arbitrary number of competing technologies $m \geq 2$, we characterize the equilibria and establish the existence of an adoption-diffused equilibrium in which all technologies coexist. For $m > 2$, the equilibrium configuration depends on the broader competitive environment: unlike the two-technology case, within-community relative adoption generally depends also on equilibrium opinions and on the other alternatives. 
    Finally, we formulate finite-horizon opinion-shaping control problems, derive necessary optimality conditions via the discrete-time Pontryagin maximum principle, and compute numerical extremals using a forward-backward sweep method.
    \end{abstract}
    
    \end{frontmatter}

\thinmuskip=0.1mu  
\medmuskip=0.1mu   
\thickmuskip=0.1mu 

\section{Introduction}
\begin{figure}[t]
    \centering
    \includegraphics[width=.6\columnwidth]{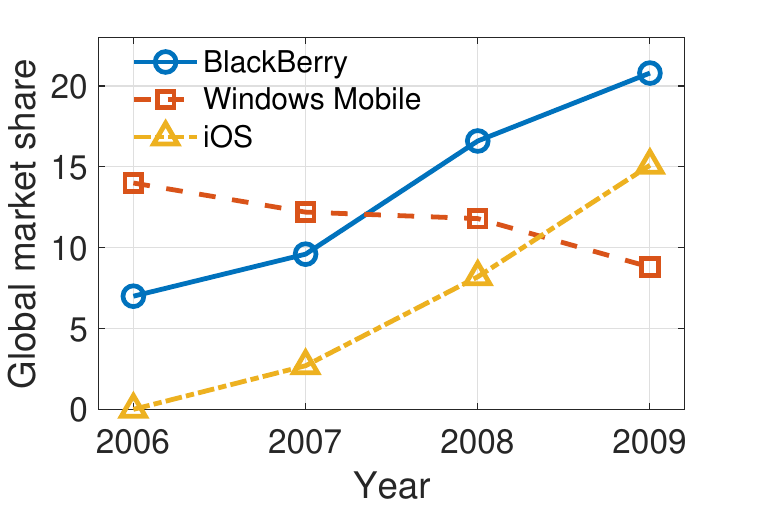}
    \caption{Global market shares of selected smartphone platforms, 2006--2009. 
    Data from \cite{kenney2011structuring}.}
    \label{fig:smartphone-data}
\end{figure}

Research on diffusion processes over networks has attracted considerable attention due to its applications ranging from epidemic spreading to the adoption of behaviors, technologies, and social norms \citep{Rogers2010, christakis2007spread, valente2005network, centola2010spread}. Adoption is often driven not only by intrinsic value of the product or behavior but also by social influence, making diffusion inherently a network-driven process emerging from local interactions.
Empirical studies of product and service diffusion have further shown that adoption decisions depend on the local network structure and on the characteristics and position of previous adopters \citep{katona2011network}.
Several modeling frameworks have been proposed to model network diffusion, including the Bass model and threshold-based approaches \citep{Bass1969, Granovetter1978, Kempe2003}, as well as epidemic-inspired models that describe adoption as a contagion process \citep{Kermack.McKendrick:1927, Pastor-Satorras2015}. In this context, SIR and SIS models, together with their extensions, have been widely used to represent spreading phenomena in social and technological systems, including reversible adoption, memory, and loss of interest \citep{goffman1964generalization, bettencourt2006power, LI20141042,daley1964epidemics}. 
However, adoption is rarely determined by exposure alone. Individual willingness to adopt may also depend on beliefs, attitudes, and perceived social norms, which naturally motivates the integration of opinion dynamics into diffusion models. Classical models such as French--DeGroot \citep{DeGroot1974} and Friedkin--Johnsen \citep{Friedkin1990} describe how opinions evolve through social interactions and persistent individual predispositions.
Building on these ideas, a growing body of work has coupled opinion evolution with an adoption dynamics resembling an epidemic one, recognizing a mutual interaction: opinions shape the willingness to adopt, while adoption itself may feed back into beliefs and attitudes \citep{granell2013dynamical, wang2019impact, lin2021discrete, she2021peak, bizyaeva2024active, Xu2024, She2022, arango2025opinion, XUAN20202582, de2024game}. A related direction is considered in \citep{ruf2019viral}, where adoption is coupled with an opinion dynamics process evolving over a signed network, thus accounting for antagonistic interactions among agents. These models show that the joint evolution of opinions and adoption can generate behaviors that contagion-based models alone cannot capture, but they typically focus on a single alternative or on restricted forms of competition.
When multiple alternatives compete for adoption, individual decisions depend not only on exposure and social influence, but also on the set of available alternatives and on the possibility of switching among them. Competition among technologies and products has long been studied in economics, where network externalities may generate strategic advantages, path dependence, and lock-in, while the introduction of new alternatives can reshape the competitive environment \citep{katz1986technology,katz1992product,david1985clio, arthur1989competing}.
Complementary game-theoretic approaches, particularly network coordination games, describe technology adoption as a strategic process in which agents benefit from coordinating their choices with those of their neighbors \citep{morris2000contagion,fagnani2017diffusion,fazeli2012game}. Competition may lead to dominance of a single technology, but long-term coexistence can also emerge in the presence of heterogeneity, switching mechanisms, or competing contagions \citep{min2018competing,liu2019analysis}. Related coexistence and competitive effects have also been observed in epidemic-inspired models of multi-virus, multi-contagion, and information spreading \citep{gracy2025modeling,sun2019model}. 
The relevance of accounting for more than two alternatives is also suggested by the evolution of real markets. Figure~\ref{fig:smartphone-data} shows that, during the expansion of iOS between 2007 and 2008, the relative ordering of Windows Mobile and BlackBerry reversed \citep{kenney2011structuring}. These aggregate data do not establish a causal effect of entry, but illustrate how competitive positions may evolve as the set of available alternatives changes.
Intervention in spreading and adoption dynamics has also been widely studied.
Empirical and model-based approaches have investigated how information provision, network targeting, and mechanisms affecting peer influence can alter diffusion, while the control literature has considered influence maximization, predictive control of technology adoption, opinion shaping, and control of competing contagions \citep{aral2011creating,Kempe2003,alutto2025predictive,
alutto2025modeling,ruf2019viral,bizyaeva2024active,liu2019analysis}.
Despite these developments, existing approaches typically capture only subsets of the mechanisms that govern competition among alternatives. In particular, competitive diffusion models generally do not account simultaneously for alternative-specific opinions, dissatisfaction with the currently adopted option, and switching, while coupled adoption-opinion control frameworks have mainly focused on a single alternative. A unified framework combining competition, endogenous opinion formation, dissatisfaction, switching, and dynamic intervention is still lacking. This leaves open the problem of understanding and controlling diffusion when several alternatives coexist and adoption decisions and opinions co-evolve over heterogeneous interaction networks.
To address this gap, we introduce the Multi-technology Adoption-opinion co-Evolution (MATE) model, a discrete-time network framework for the simultaneous diffusion of $m\geq2$ competing technologies over two heterogeneous interaction layers. The model tracks susceptible, adopter, and dissatisfied individuals together with technology-specific opinions, allowing adoption, dissatisfaction, switching, and social influence to co-evolve within a unified framework. Although we formulate the model in terms of competing technologies, the same structure applies more generally to products, services, platforms, or other competing alternatives. The present work extends the preliminary two-alternative formulation in \citep{alutto2026coupled} to an arbitrary number of competing alternatives. For $m>2$, equilibrium outcomes depend on the full set of available alternatives, leading to a qualitatively richer competitive structure than in the two-alternative case.\\
The contributions of this paper are threefold. First, we generalize the MATE framework to $m\geq2$ competing alternatives (Section~\ref{sec:model}) and establish the well-posedness of its dynamics. Second, we characterize the \emph{adoption-diffused equilibria} (Section~\ref{sec:equilibria_general}), 
establishing existence under strictly positive dissatisfaction rates, and providing sufficient conditions for uniqueness and global convergence from non-adoption-free initial states. We further identify features specific to the multi-alternative setting: the cancellation underlying the two-alternative adoption ratio is lost, pairwise adoption depends on the broader competitive environment, and uniqueness of the adoption-diffused equilibrium is no longer guaranteed under the standing assumptions. Numerical examples additionally illustrate the coexistence of multiple equilibria and entry-induced reversals between incumbent alternatives.
Third, we formulate two finite-horizon \emph{optimal opinion-shaping control} problems, derive necessary conditions via the discrete-time Pontryagin maximum principle and compute numerical extremals using a forward-backward sweep algorithm (Section~\ref{sec:opinion-control}).

\subsection{Notation}
We denote by $\mathbb{R}$ and $\mathbb{R}_{+}$ the sets of real and nonnegative real numbers, respectively, and by $\mathbb{R}_{+}^{n\times n}$ the set of $n\times n$ real matrices with nonnegative entries.
The all-ones and all-zeros vectors are denoted by $\boldsymbol{1}$ and $\boldsymbol{0}$, respectively, and the identity matrix by $I$.
The transpose of a matrix $A$ is denoted by $A^{\top}$.
For $x\in\mathbb{R}^n$, we define $\|x\|_{1}=\sum_i |x_i|$ and $\|x\|_{\infty}=\max_i |x_i|$ as its $\ell_{1}$ and $\ell_{\infty}$ norms, and we write $\diag(x)$ for the diagonal matrix with diagonal entries given by $x$.
For an irreducible matrix $A\in\mathbb{R}_{+}^{n\times n}$ (whose associated directed graph is strongly connected), we denote by $\rho(A)$ the spectral radius of $A$.
Vector inequalities $x\le y$ for $x,y\in\mathbb{R}^n$ are understood component-wise.

\section{Model Description}\label{sec:model}
\begin{figure}[t]
\centering
\begin{tikzpicture}[
    >=Stealth,
    font=\small,
    node distance=1.5cm and 1.5cm,
    comp/.style={
        draw,
        circle,
        minimum size=8mm,
        inner sep=0.8pt,
        line width=0.8pt
    },
    agg/.style={
        draw,
        rounded corners=2pt,
        minimum width=18mm,
        minimum height=5mm,
        inner sep=1.5pt,
        align=center,
        line width=0.8pt
    },
    flow/.style={->, line width=0.9pt},
    switchflow/.style={->, dashed, line width=0.9pt}
]
\node[comp] (S) at (0,0) {$S$};
\node[comp] (A1) at (3.5,1.8) {$A^{[1]}$};
\node[comp] (D1) at (7.5,1.8) {$D^{[1]}$};
\node[agg]  (U1) at (5,3.1) {$\sum_{\ell\neq 1} D^{[\ell]}$};
\node[comp] (A2) at (3.5,0) {$A^{[2]}$};
\node[comp] (D2) at (7.5,0) {$D^{[2]}$};
\node[agg]  (U2) at (5,1.3) {$\sum_{\ell\neq 2} D^{[\ell]}$};
\node at (3.5,-0.9) {$\vdots$};
\node at (7.5,-0.9) {$\vdots$};
\node at (4,-0.1) {};
\node[comp] (Am) at (3.5,-1.8) {$A^{[m]}$};
\node[comp] (Dm) at (7.5,-1.8) {$D^{[m]}$};
\node[agg]  (Um) at (5,-0.5) {$\sum_{\ell\neq m} D^{[\ell]}$};
\draw[flow] (S) -- node[above left, pos=0.55] {$\beta^{[1]}x^{[1]}$} (A1);
\draw[flow] (S) -- node[above, pos=0.55] {$\beta^{[2]}x^{[2]}$} (A2);
\draw[flow] (S) -- node[below left, pos=0.55] {$\beta^{[m]}x^{[m]}$} (Am);

\draw[flow] (A1) -- node[above right, pos=0.5] {$\delta^{[1]}$} (D1);
\draw[flow] (A2) -- node[above right, pos=0.5] {$\delta^{[2]}$} (D2);
\draw[flow] (Am) -- node[above right, pos=0.5] {$\delta^{[m]}$} (Dm);

\draw[switchflow] (U1) -- node[right, pos=0.55] {$\gamma^{[1]}x^{[1]}$} (A1);
\draw[switchflow] (U2) -- node[right, pos=0.55] {$\gamma^{[2]}x^{[2]}$} (A2);
\draw[switchflow] (Um) -- node[right, pos=0.55] {$\gamma^{[m]}x^{[m]}$} (Am);
\end{tikzpicture}
\caption{Schematic transition diagram of the multi-technology adoption-opinion model. Susceptible individuals may adopt technology $k$ and enter $A^{[k]}$, adopters may become dissatisfied and move to $D^{[k]}$ at rate $\delta^{[k]}$, and dissatisfied individuals may switch to competing alternatives.}
\label{fig:transitions_m}
\end{figure}
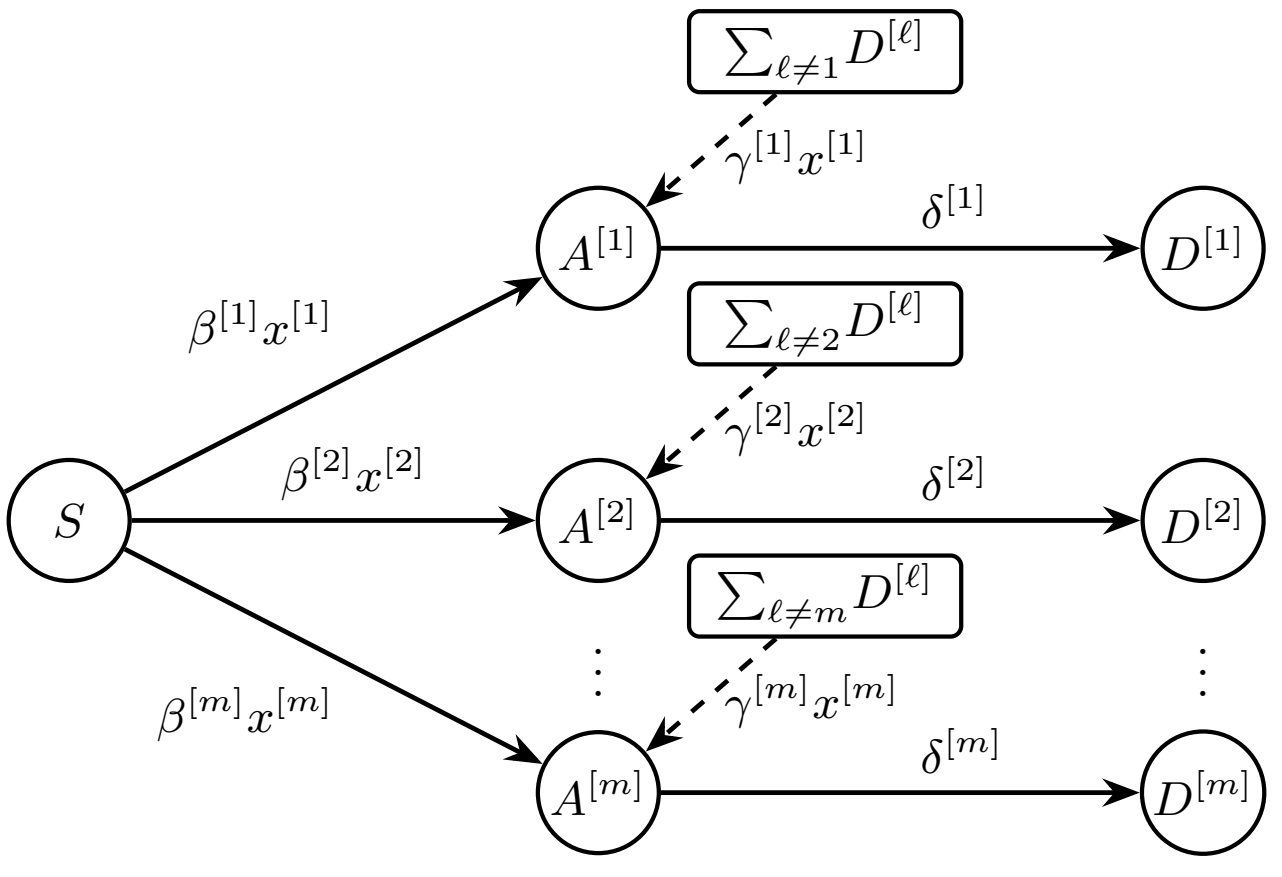

We consider a population divided into $n$ communities, where each community represents a subpopulation of indistinguishable individuals that can interact and adopt one among $m\ge 2$ competing technologies or products. Let $\Kset:=\{1,\dots,m\}$ denote the alternative set. Within each community $i$ belonging to  $\V:= \{ 1,\dots,n\}$, individuals are classified into the following compartments at each time instant $t\geq0$:
\begin{itemize}
    \item $s_i(t)$: fraction of susceptibles to all alternatives, who have not adopted any technology,
    \item $a_i^{[k]}(t)$: fraction of adopters of technology $k\in\Kset$,
    \item $d_i^{[k]}(t)$: fraction of dissatisfied of technology $k\in\Kset$.
\end{itemize}
The compartmental structure describing the adoption process is illustrated in Figure~\ref{fig:transitions_m}, where arrows indicate the possible transitions between the susceptible ($S$), adopter ($A^{[k]}$), and dissatisfied ($D^{[k]}$) states. Susceptible individuals may adopt technology $k$, upon exposure to adopters, while adopters may become dissatisfied due to a negative experience at a technology-dependent rate. Dissatisfied individuals can switch to an alternative depending on their opinions.
The interactions among communities are captured by a weighted directed network $\G = (\V, \mc E, W)$, where $\mc E \subseteq \V \times \V$ and $W\in\R_+^{n\times n}$ is a nonnegative \emph{physical interaction matrix}. Each entry $W_{ij}$ quantifies the strength of influence that community $j$ exerts on community $i$ in terms of adoption exposure.

As mentioned above, in addition to physical interactions, adoption is driven by individual beliefs. To capture this effect, each community $i$ is associated with an \emph{opinion variable} $x_i^{[k]} \in [0,1]$ reflecting the average interest toward adopting technology $k$. We further denote by $p_i^{[k]}\in[0,1]$ the fixed predisposition of community $i$ toward alternative $k$. Opinions evolve over a social network $\mc{\tilde G} = (\mc V, \mc{\tilde E}, \mc{\tilde{W}})$, where $\tilde{\mc E} \subseteq \mc V \times \mc V$ is the set of directed links, and $\tilde{W} \in \R_+^{n \times n}$ is a nonnegative \emph{social interaction matrix}. Each entry $\tilde W_{ij}$ quantifies the strength of influence that community $j$ exerts on community $i$ in shaping its opinion toward the available alternatives.
Favorable opinions increase the effective adoption rate and the propensity to switch to the corresponding alternative. These effective rates are modeled as $\beta_i^{[k]}x_i^{[k]}$ and $\gamma_i^{[k]}x_i^{[k]}$, respectively, with  $\beta_i^{[k]}, \gamma_i^{[k]} \in (0,1)$. Dissatisfaction instead occurs at a technology-dependent rate $\delta_i^{[k]}\in[0,1]$, independently of opinions, reflecting intrinsic quality or user-experience factors.
For each $i\in\V$, $k\in\Kset$, the dynamics is given by
\begin{subequations}\label{eq:model_general}
\allowdisplaybreaks
\begin{align}
s_i(t+1)=& s_i(t) \!-\! s_i(t)\! \sum_{k\in\Kset} \beta_i^{[k]} x_i^{[k]}(t) \!\sum_{j\in \mc V}\! W_{ij} a_j^{[k]}(t), \label{eq:gen_s}\\[-0.2ex]
a_i^{[k]}(t+1)=& a_i^{[k]}(t) + \beta_i^{[k]} x_i^{[k]}(t) s_i(t)\! \sum_{j\in \mc V} W_{ij} a_j^{[k]}(t) \nonumber\\[-0.5ex]
&\ - \delta_i^{[k]} a_i^{[k]}(t) + \gamma_i^{[k]} x_i^{[k]}(t)\!\!\! \sum_{\ell\in\Kset\setminus\{k\}} \! \!\! d_i^{[\ell]}(t),\label{eq:gen_a}\\[-0.2ex]
d_i^{[k]}(t+1)=& d_i^{[k]}\!\!(t) - d_i^{[k]}\!\!(t)\mspace{-15mu} \sum_{\ell\in\Kset\setminus\{k\}}\!\! \! \gamma_i^{[\ell]} x_i^{[\ell]}(t) + \delta_i^{[k]}\! a_i^{[k]}\!(t),\label{eq:gen_d}\\\nonumber
x_i^{[k]}(t+1)=& (1 - \lambda_i^{[k]} - \xi_i^{[k]}) p_i^{[k]}  \\
&+ \lambda_i^{[k]} \sum_{j\in \mc V} \widetilde{W}_{ij} x_j^{[k]}(t)\!+\! \xi_i^{[k]} \sum_{j\in \mc V} W_{ij} a_j^{[k]}(t),\label{eq:gen_x}
\end{align}
\end{subequations}
where $\lambda_i^{[k]},\xi_i^{[k]}\ge 0$ satisfy $\lambda_i^{[k]}+\xi_i^{[k]} < 1$ for all $k\in\Kset$. The proposed opinion dynamics \eqref{eq:gen_x} extends the classical Friedkin--Johnsen model \cite{Friedkin1990} by incorporating a feedback term from the adoption layer. Specifically, opinions towards a technology are shaped by three factors: the fixed predisposition of each community, social interactions over the opinion network, and the observed adoption levels in the physical layer. Hence, $\lambda_i^{[k]}$ quantifies the susceptibility to social influence and $\xi_i^{[k]}$ captures the strength of adoption-driven feedback for technology $k$ for community $i$. Note that the variables $x_i^{[k]}$ represent alternative-specific levels of interest rather than mutually exclusive choice probabilities.

We refer to \eqref{eq:model_general} as the \textit{Multi-technology AdopTion-opinion co-Evolution} (MATE) model. This framework extends the adoption--opinion models proposed in \citep{alutto2025modeling,alutto2025predictive}, which focus on the diffusion of a single technology, and generalizes the preliminary two-technology setting studied in \citep{alutto2026coupled} to the case of $m\geq 2$ competing alternatives.

\subsection{Preliminary assumptions and properties}
We introduce the standing assumptions used throughout the analysis.
\begin{assumption}\label{ass:main}
\begin{itemize}
    \item[(i)] The matrices $W$ and $\widetilde{W}$ are row-stochastic, and $W$ is irreducible (i.e., the associated directed graph is strongly connected).
    \item[(ii)] For all $k\in\Kset$, the predisposition satisfies $p^{[k]}\!>\!\boldsymbol{0}$.
    \item[(iii)] For each $i\in \mc V$, it holds that $\sum_{k\in\Kset}\beta_i^{[k]}\le 1$ and $\sum_{k\in\Kset}\gamma_i^{[k]}\le 1$.
\end{itemize}
\end{assumption}

Assumption \ref{ass:main}(i) implies that the interaction terms between communities are convex combinations of the fractions of adopters at neighboring communities. This guarantees that both the overall exposure term $\sum_j W_{ij}a_j^{[k]}$ and the social imitation term $\sum_j \widetilde W_{ij}x_j^{[k]}$ remain in $[0,1]$. Moreover, the irreducibility of $W$ excludes the presence of disconnected components in the physical network.
Assumption \ref{ass:main}(ii) ensures a strictly positive baseline predisposition toward
each alternative, while 
Assumption \ref{ass:main}(iii) imposes natural bounds on the transition rates, so that the total flow from each compartment is compatible with the interpretation of the state variables as fractions. 

Let $a^{[k]}(t),d^{[k]}(t),x^{[k]}(t)\in\R^n$ collect community states for technology $k\in\Kset$ and define the following diagonal matrices: 
$B^{[k]}:=\diag(\beta^{[k]})$, $\Gamma^{[k]}:=\diag(\gamma^{[k]})$, $\Delta^{[k]}:=\diag(\delta^{[k]})$, $\Lambda^{[k]}:=\diag(\lambda^{[k]})$, $\Xi^{[k]}:=\diag(\xi^{[k]})$. We define the admissible set of variables 
\begin{multline*}
\X:=\Big\{(s,(a^{[k]})_{k},(d^{[k]})_{k},(x^{[k]})_{k})\in[0,1]^{(3m+1)n}:\\[-0.7ex]
s+\sum\nolimits_{k\in\Kset}(a^{[k]}+d^{[k]})=\boldsymbol{1}\Big\}.
\end{multline*}
For simplicity of notation, from now on, the state of the MATE model \eqref{eq:model_general} at time $t \geq 0$ is denoted as
\be \label{eq:state} y(t):=(s(t),(a^{[k]})_{k}(t),(d^{[k]})_{k}(t),(x^{[k]})_{k}(t)).\ee
\begin{proposition}\label{prop:invariant_general}
    Consider the MATE model \eqref{eq:model_general} under Assumption~\ref{ass:main}. If the initial condition belongs to $\X$, then the trajectory remains in $\X$ for all $t\ge 0$.
\end{proposition}

\begin{corollary}\label{cor:monotone_s}
    Consider the MATE model \eqref{eq:model_general} under Assumption~\ref{ass:main}. If the initial condition belongs to $\X$, then the following hold:
    \begin{itemize}
        \item[(i)] $s(t+1)\le s(t)$ for all $t\ge 0$.
        \item[(ii)] $x^{[k]}(t) > \boldsymbol{0}$ for all $t \geq 1$, $k \in \Kset$.
    \end{itemize}
\end{corollary}
In particular, Corollary~\ref{cor:monotone_s} establishes nonincreasing monotonicity of the susceptibles.
The proofs of Proposition~\ref{prop:invariant_general} and Corollary~\ref{cor:monotone_s} are in Appendix~\ref{apx:proof_sec_2}.

\section{Equilibrium points and stability analysis}\label{sec:equilibria_general}
In this section, we analyze the equilibrium points of the adoption–opinion model \eqref{eq:model_general} and study their stability properties. We first analyze the \emph{adoption-free equilibrium} (AFE) (Section~\ref{sec:adoption-free-eq}), namely the equilibrium configuration in which no individuals adopt any technology and the whole population remains susceptible, and then the \emph{adoption-diffused equilibrium} (ADE) (Section~\ref{sec:adoption-diffused-eq}), characterized by a positive fraction of adopters and dissatisfied individuals across communities. Finally, we discuss the special case of $m=2$ alternatives (Section~\ref{sec:two_tech}) and show that the within-community equilibrium adoption ratio depends only on dissatisfaction (user-experience) parameters. The proofs of all results in this section can be found in Appendix \ref{apx:proof_sec_3}.  

\subsection{Adoption-free equilibrium}\label{sec:adoption-free-eq}
\begin{proposition}\label{prop:afe}
    The MATE model \eqref{eq:model_general} under Assumption~\ref{ass:main} admits the AFE, defined as follows
    \be\label{eq:y_afe} y_{AF}:=\big(\boldsymbol{1},( \boldsymbol{0})_{k\in\Kset},(\boldsymbol{0})_{k\in\Kset},(x_{AF}^{[k]})_{k\in\Kset}\big),    \ee
    where, for each $k\in\Kset$,
    \be\label{eq:x_afe} x_{AF}^{[k]}:= (I-\Lambda^{[k]} \widetilde{W})^{-1}\big((I-\Lambda^{[k]}-\Xi^{[k]})p^{[k]}\big). 
    \ee
    Moreover, $y_{AF}$ is not attractive and, consequently, is not asymptotically stable.
\end{proposition}
\begin{rem}
    Proposition \ref{prop:afe} shows that $y_{AF}$ is not attractive: arbitrarily small perturbations introducing a positive fraction of adopters do not converge back to $y_{AF}$. Therefore, $y_{AF}$ cannot be asymptotically stable. 
    A stronger claim of instability would require showing that trajectories starting arbitrarily close to $y_{AF}$ eventually leave a fixed neighborhood of $y_{AF}$, which is not established here.
\end{rem}


\subsection{Adoption-diffused equilibrium}\label{sec:adoption-diffused-eq}
This section characterizes the equilibrium structure of the MATE model beyond the adoption-free state. We first establish the existence of an ADE, in which the susceptible population is depleted throughout the network, and characterize its structure. We then rule out partial-adoption equilibria and provide sufficient conditions for the uniqueness and convergence to the ADE.

Before showing the first result, we denote any equilibrium with $y^{\star}$ and define for each community $i$ these aggregate quantities:
\be \label{eq:def-bar-dx}
\bar d_i^{\star}:=\sum\nolimits_{k\in\Kset} d_i^{[k]{\star}},\quad 
\bar x_i^{\star}:=\sum\nolimits_{k\in\Kset}\gamma_i^{[k]} x_i^{[k]{\star}}.
\ee
We start proving the existence of an equilibrium with $s^{\star}=\boldsymbol{0}$ and strictly positive adoption/dissatisfaction components.
\begin{theorem}\label{thm:exist_general}
    Consider the MATE model \eqref{eq:model_general} under Assumption~\ref{ass:main}. Assume $\delta_i^{[k]} >0$ for all $i \in \mc V$ and $k \in \Kset$. Then, there exists an ADE, defined as follows 
    \be \label{eq:ade}y^{\star}:=(s^{\star}, (a^{[k]{\star}})_{k \in \Kset}, (d^{[k]{\star}})_{k \in \Kset}, (x^{[k]{\star}})_{k \in \Kset})\ee 
    such that $s^{\star}=\boldsymbol{0}$ and $a^{[k]{\star}}>\boldsymbol{0}$, $d^{[k]{\star}}>\boldsymbol{0}$ for all $k\in\Kset$, while the opinion vector satisfies, for each $k$,
    \be\label{eq:x_star_general} \! x^{[k]{\star}}\! = (I \!-\! \Lambda^{[k]}\widetilde{W})^{-1}\! \big(\!\!(I\!-\!\Lambda^{[k]}\!-\Xi^{[k]})p^{[k]}+\Xi^{[k]}W a^{[k]{\star}}\!\!\big).\ee
    Moreover, the following structural identities hold:
    \be \label{eq:a_formula}
        d_i^{[k]\star}\!\! = \frac{\gamma_i^{[k]} x_i^{[k]\star}}{\bar x_i^\star} \bar d_i^\star,\mspace{12mu}
        a_i^{[k]\star}\!\! = \frac{\gamma_i^{[k]} x_i^{[k]\star}}{\delta_i^{[k]}} \bar d_i^\star \Bigg(\!\! 1-\frac{\gamma_i^{[k]} x_i^{[k]\star}}{\bar x_i^\star}\!\!\Bigg),  
    \ee
    where $\bar d_i^{\star}$ and $\bar x_i^{\star}$ are defined in \eqref{eq:def-bar-dx}.
\end{theorem}

\begin{rem}
    Eq. \eqref{eq:a_formula} shows an inverse dependence of the fraction of adopters $a_i^{[k]\star}$ on the dissatisfaction rate $\delta_i^{[k]}$ when the remaining equilibrium quantities are held fixed, which is consistent with the interpretation that higher dissatisfaction accelerates the abandonment of adoption.
\end{rem}
\begin{rem}
    The assumption $\delta_i^{[k]}>0$ for all $i\in\mc V$ and $k\in\Kset$ is essential for a strictly positive ADE. At any equilibrium with $s_i^\star=0$, if some dissatisfaction rates vanish at node $i$, equilibrium adoption can be supported only on the corresponding alternatives, while all dissatisfaction components vanish. In particular, if $\delta_i^{[k_0]}=0$ for a unique $k_0$, then $a_i^{[k_0]\star}=1$, yielding a local monopoly.
\end{rem}

The previous theorem characterizes equilibria with no susceptibles. The following proposition shows that, under Assumption~\ref{ass:main}, an equilibrium cannot exhibit adoption in only a subset of communities.
\begin{proposition}\label{prop:no_partial_general}
    Consider the MATE model \eqref{eq:model_general} under Assumption~\ref{ass:main}. Let $y^{\star}$ be an equilibrium for \eqref{eq:model_general}. If there exists at least one node $i \in \mc V$ such that $s_i^{\star}=0$, then $s^{\star}=\boldsymbol{0}$.
\end{proposition}

\begin{corollary}\label{cor:equilibrium_dichotomy}
Consider the MATE model~\eqref{eq:model_general} under
Assumption~\ref{ass:main}, and assume that $\delta_i^{[k]}>0$ for all $i\in\V$ and $k\in\Kset$. Then every equilibrium is either the AFE or an ADE.
\end{corollary}


The next results establish sufficient conditions for the uniqueness and convergence to the ADE.
\begin{theorem}\label{thm:unique}
    Consider the MATE model \eqref{eq:model_general} under Assumption~\ref{ass:main}. Assume that $\delta_i^{[k]}>0$ for all $i\in\V$ and $k\in\Kset$. If
    \be \label{eq:hyp-unique} 
    \max_{k\in\Kset}\Bigg(\frac{\max_i \xi_i^{[k]}}{1- \max_i \lambda_i^{[k]}} \max_i \frac{1}{\delta_i^{[k]}}\Bigg(\! 1\!+\! \frac{1}{ 2\min_{r}\delta_i^{[r]}}\Bigg)\Bigg)<1. \ee
    then the ADE is unique. 
\end{theorem}
\begin{rem}
    The sufficient condition in \eqref{eq:hyp-unique} ensures that the feedback between opinions and adoption is sufficiently weak. The first factor bounds the sensitivity of the opinions at equilibrium to adoption, while the second bounds the sensitivity of the adoption at equilibrium to opinions. 
\end{rem}

Before the next result, define the following for all $i$,
\begin{subequations}
\be \label{eq:params-def}
\begin{aligned}    &\delta_{i,\min}:=\min\nolimits_k\delta_i^{[k]},\quad \gamma_{i,\min}:=\min\nolimits_k\gamma_i^{[k]}, \\ 
&\underline x_i:=\min\nolimits_k(1-\lambda_i^{[k]}-\xi_i^{[k]})p_i^{[k]}>0, \end{aligned} \ee
and
\be \label{eq:vareps}\varepsilon_i:=\underline x_i\delta_{i,\min}^2\gamma_{i,\min}^3>0. \ee
\end{subequations}
\begin{theorem}\label{thm:convergence}
    Consider the MATE model \eqref{eq:model_general} under Assumption~\ref{ass:main}. 
    Assume that $\delta_i^{[k]}>0$ for all $i\in\V$ and $k\in\Kset$. 
    If 
    \be \label{eq:hyp-convergence} \frac{6}{\min_i \varepsilon_i}\mspace{3mu} \max_{k\in\Kset}\frac{\max_i\xi_i^{[k]}}{1-\max_i\lambda_i^{[k]}} <1,\ee
    then every trajectory with $s(0)\neq \boldsymbol{1}$ converges to the ADE $y^\star$, namely $y(t)\to y^\star$ as $t\to\infty$.
\end{theorem}
\begin{rem}
    The sufficient condition~\eqref{eq:hyp-convergence} requires the feedback between opinion variations and adopter-dissatisfied dynamics to be sufficiently weak.
\end{rem}

\begin{figure}
    \hspace{-0.3cm}
    \subfloat[]{\includegraphics[width=0.525\linewidth]{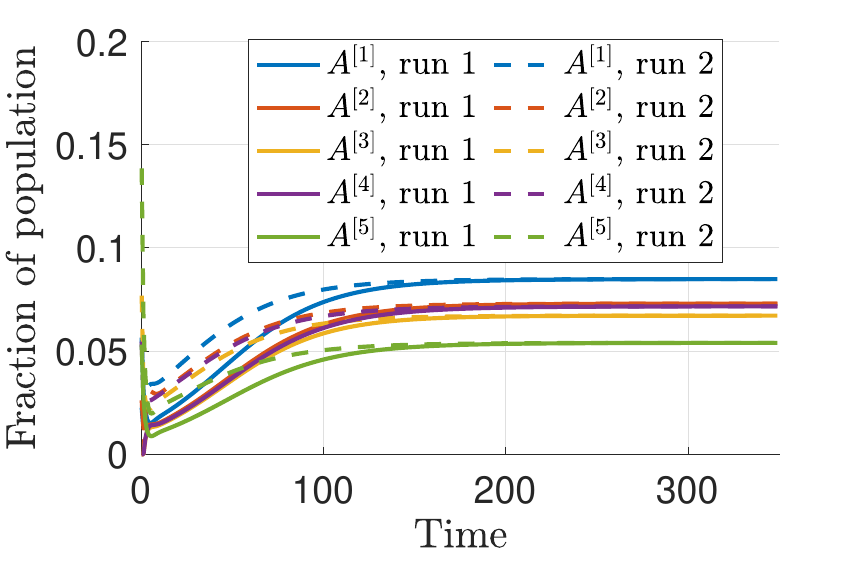}}
    \hspace{-0.3cm}
    \subfloat[]{\includegraphics[width=0.525\linewidth]{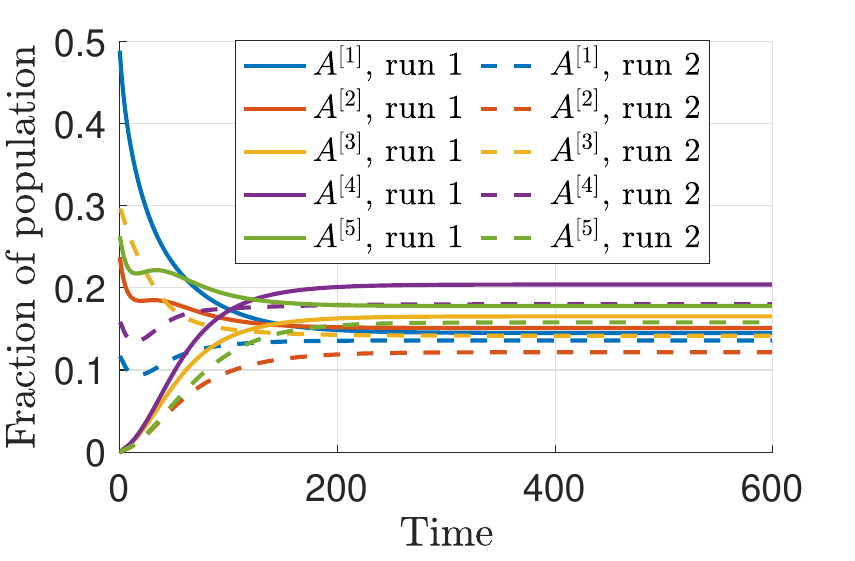}}
    \caption{Time evolution of the average population fractions with $m=5$ alternatives. 
    (a) Condition~\eqref{eq:hyp-unique} holds and two trajectories with distinct initial conditions converge to the same ADE. (b) Condition~(9) is violated and two different trajectories converge to distinct ADEs. }
    \label{fig:uniqueness}
\end{figure}

\begin{example}
    To illustrate the uniqueness result of Theorem~\ref{thm:unique}, we consider a network with $n=30$ communities and $m=5$ competing alternatives, with parameters randomly selected within the ranges allowed by Assumption~\ref{ass:main}. 
    Let $A^{[k]}(t):=1/n \boldsymbol{1}^\top a^{[k]}(t)$ denote the mean adoption of technology $k$. 
    Fig.~\ref{fig:uniqueness}(a) considers a parameter regime satisfying condition~\eqref{eq:hyp-unique}. Trajectories initialized from different adoption profiles approach the same equilibrium, consistently with the uniqueness of the ADE.
    In contrast, Fig.~\ref{fig:uniqueness}(b) considers a parameter regime in which the uniqueness condition \eqref{eq:hyp-unique} is violated. In this case, two different initial adoption profiles converge numerically to distinct equilibria, showing that multiple ADEs and path-dependent long-term outcomes may arise when the sufficient uniqueness condition is not satisfied.
\end{example}
Although convergence is theoretically guaranteed only under the sufficient condition \eqref{eq:hyp-convergence}, numerical experiments suggest that convergence may also occur beyond this regime.
\subsection{Special Case: Two Technologies}\label{sec:two_tech}  
We now specialize the analysis to $\Kset=\{1,2\}$, the $m=2$ technologies case introduced in \cite{alutto2026coupled}, and highlight analogies and differences with the general multi-technology case, focus of the present work.

\begin{proposition}\label{thm:ratio_m2}
Consider the MATE model \eqref{eq:model_general} with $m=2$ under Assumption~\ref{ass:main} and let $\delta_i^{[k]}>0$ for all $i\in\V$ and $k\in\{1,2\}$. Then, there exists a unique ADE $y^\star$ satisfying $s^\star=\boldsymbol{0}$ and, for all $i\in\V$,
\begin{equation}\label{eq:ratio_m2}
\frac{a_i^{[2]\star}}{a_i^{[1]\star}}=\frac{\delta_i^{[1]}}{\delta_i^{[2]}},
\qquad \frac{d_i^{[2]\star}}{d_i^{[1]\star}}=\frac{\gamma_i^{[2]}x_i^{[2]\star}}{\gamma_i^{[1]}x_i^{[1]\star}}.
\end{equation}
\end{proposition}
The result follows from Theorem~\ref{thm:exist_general} specialized to $m=2$, together with the uniqueness result in \cite[Theorem~5]{alutto2026coupled}; hence, the proof is omitted.

\begin{rem}
    For $m=2$, the within-community equilibrium adoption ratio in \eqref{eq:ratio_m2} depends only on dissatisfaction (user-experience) parameters and is, therefore, independent of the equilibrium opinions and switching parameters. This follows from a cancellation that is specific to the two-technology case.
    By contrast, for $m>2$, this simplification no longer holds: the factor $\big(1-{\gamma_i^{[k]} x_i^{[k]{\star}}}/{\bar x_i^{\star}}\big)$ in \eqref{eq:a_formula} prevents such cancellation, so the equilibrium shares generally depend also on the equilibrium opinions and the overall attractiveness of switching. On the other hand, the ratio of dissatisfied individuals is still governed by the relative attractiveness of switching, namely by the quantities $\gamma_i^{[k]}x_i^{[k]\star}$.
\end{rem}
The extension from two to multiple competing technologies substantially changes the equilibrium properties. For $m=2$, uniqueness of the ADE follows without the additional contraction condition required in the multi-alternative case. 

\begin{figure}
    \centering
    \includegraphics[width=0.65\linewidth]{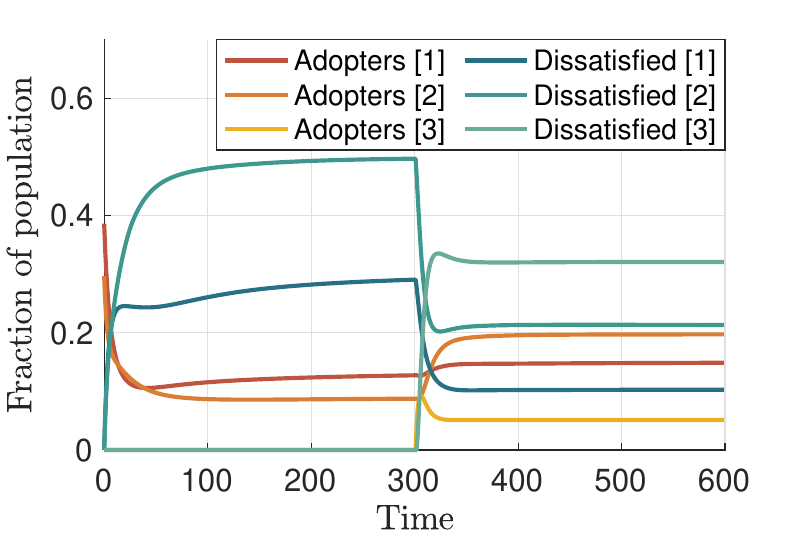}
    \caption{Effect of technology entry on the adoption dynamics. The introduction of technology~3 changes the relative ordering of technologies~1 and~2, while the entrant remains non-dominant.}
    \label{fig:kingmaker}
\end{figure}

\begin{example}
We illustrate a competitive effect specific to $m>2$ by first evolving the MATE model with two incumbent technologies and introducing a third at time $t_e$, while leaving the incumbents unchanged. The two regimes correspond to the MATE model with $m=2$ and $m=3$, respectively. At entry, the incumbent states are unchanged and the entrant is initialized with $a^{[3]}(t_e)=d^{[3]}(t_e)=\boldsymbol{0}$ and $x^{[3]}(t_e)=p^{[3]}$. As shown in Fig.~\ref{fig:kingmaker}, technology~1 initially has larger adoption than technology~2, whereas after entry their ordering is reversed, although technology~3 remains non-dominant. Thus, the entrant reshapes the competition through opinion-dependent switching.
\end{example}

\section{Opinion shaping control}\label{sec:opinion-control}
Having characterized the equilibrium and convergence properties of the MATE dynamics~\eqref{eq:model_general}, we now investigate how external interventions can steer technology adoption through the opinion layer. We consider interventions that modify the baseline predisposition toward a technology without directly altering the adoption, dissatisfaction, or switching rates. Such interventions may represent information campaigns, educational initiatives, advertising, or targeted communication aimed at improving individuals' perceptions of a technology rather than directly enforcing adoption. Empirical studies suggest that these interventions can significantly affect attitudes and adoption intentions \citep{li2020analysis}.
We model the control as a shift of the fixed predisposition $p^{[k]}$ entering the Friedkin--Johnsen component of the opinion dynamics, while social influence and adoption feedback continue to evolve endogenously. We consider two complementary intervention objectives: promoting cumulative adoption across all alternatives and maximizing the terminal adoption of a selected target technology under a prescribed intervention budget.


\subsection{Optimal control problem definition}
For each community $i\in\mc V$ and technology $k\in\Kset$, the control modifies the predisposition according to $ p_{U,i}^{[k]}(t):= p_i^{[k]}(1-u_i^{[k]}(t))+u_i^{[k]}(t)$, with $u_i^{[k]}(t)\in[0,1]$. Thus, $u_i^{[k]}(t)=0$ leaves the predisposition unchanged, whereas $u_i^{[k]}(t)=1$ shifts it to its maximal value. The controlled opinion dynamics becomes
\begin{multline}\label{eq:dynamic_controlled_x}
    x_i^{[k]}(t+1) =(1-\lambda_i^{[k]}-\xi_i^{[k]})p_{U,i}^{[k]}(t)\\
 +\lambda_i^{[k]}\sum\nolimits_{j\in \mc V}\widetilde W_{ij}x_j^{[k]}(t)+\xi_i^{[k]}\sum\nolimits_{j\in \mc V} W_{ij}a_j^{[k]}(t),
\end{multline}
while the variables $s_i(t)$, $a_i^{[k]}(t)$, and $d_i^{[k]}(t)$ evolve according to the original adoption-dissatisfaction dynamics \eqref{eq:gen_s}--\eqref{eq:gen_d}. 
The pointwise control set is $U:=[0,1]^{n|\Kset|}\subset\mathbb R^{n|\Kset|}$ and the set of admissible control sequences is $\mc U:=\{u(\cdot)\!=\!(u(0),\dots,u(T-1))\!:u(t)\in U,\ t\in[0,T-1]\}$, where $u(t)$ collects the control inputs for each technology in $\Kset$ and each community in $\mc V$.
For compactness, we write the controlled dynamics, given by \eqref{eq:dynamic_controlled_x} and \eqref{eq:gen_s}--\eqref{eq:gen_d}, as
\be y(t+1)=f(y(t),u(t)),\qquad y(0)=y_0, \label{eq:compact_system} \ee
where $y(t)$ collects all variables $s_i(t)$, $a_i^{[k]}(t)$, $d_i^{[k]}(t)$, and $x_i^{[k]}(t)$, as defined in \eqref{eq:state}. Since $p_{U,i}^{[k]}(t)\in[p_i^{[k]},1]$, the invariance argument of Proposition \ref{prop:invariant_general} extends directly to the controlled dynamics \eqref{eq:compact_system}.

To quantify the intervention effort, we assign a positive cost to each control component. Let $Q=\operatorname{diag}(q_i^{[k]})\in\mathbb R^{nm \times nm}$ be positive definite. Then $u(t)^\top Q u(t)=\sum\nolimits_{k\in\Kset}\sum\nolimits_{i \in \mc V} q_i^{[k]}\left(u_i^{[k]}(t)\right)^2$, 
where $q_i^{[k]}>0$ weights the cost of controlling the opinion of node $i$ toward technology $k$.
For the targeted intervention toward technology $k_0$, we restrict the admissible controls to $U_{k_0}:= \Big\{ u\in [0,1]^{nm}: u_i^{[k]}=0 \ \forall i\in\mc V,\ k\neq k_0 \Big\}$, and denote by $\mc U_{k_0}$ the corresponding set of admissible control sequences over the time horizon.
We can now formulate the two finite-horizon optimal control problems.
\begin{problem}[Cumulative adoption maximization] \label{prob:cumulative_adoption}
Given weights $r_1^{[k]}>0$, representing the relative importance assigned to adoption of technology $k$, find an admissible control $u^*\in\mc U$ maximizing
\beu J_1(u)=\sum\nolimits_{t=0}^{T-1}\Big(\sum\nolimits_{k\in\Kset} r_1^{[k]}\boldsymbol 1^\top a^{[k]}(t) -u(t)^\top Q u(t)\Big), \label{eq:objective_cumulative} \eeu
subject to the controlled dynamics \eqref{eq:compact_system}.
\end{problem}
\begin{problem}[Targeted adoption maximization] \label{prob:terminal_adoption}
Given a target technology $k_0\in\Kset$, and a budget $B>0$, find an admissible control $u^*\in\mc U_{k_0}$ maximizing
\beu J_2(u)=\boldsymbol 1^\top a^{[k_0]}(T), \label{eq:objective_terminal} \eeu
subject to the controlled dynamics \eqref{eq:compact_system} and the budget constraint
\be \label{eq:budget-constraint} \sum\nolimits_{t=0}^{T-1}u(t)^\top Q u(t) \leq B. \ee
\end{problem}

Both problems can be written in the common form
$J(u) =\sum_{t=0}^{T-1} \ell(y(t),u(t)) + \Phi(y(T))$.
For Problem~\ref{prob:cumulative_adoption}, we have $\ell_1(y(t),u(t)) :=\sum_{k\in\Kset} r_1^{[k]}\boldsymbol 1^\top a^{[k]}(t) - u(t)^\top Q u(t)$, $ \Phi_1(y(T)) :=0$.
For Problem~\ref{prob:terminal_adoption}, we have $\ell_2(y(t),u(t)) := 0$, $\Phi_2(y(T)) := \boldsymbol 1^\top a^{[k_0]}(T) $. 
\begin{rem}\label{rem:control}
    The control affects adoption with a two-step delay, since $u(t)$ modifies $x(t+1)$, which in turn affects $a(t+2)$. Consequently, in Problem~\ref{prob:cumulative_adoption} the last two control actions do not affect the adoption reward and are therefore optimally zero. In Problem~\ref{prob:terminal_adoption}, the last control action does not affect the terminal adoption level and can be set to zero without loss of optimality.
\end{rem}

\subsection{Existence of optimal controls}
We first show that both finite-horizon formulations admit an optimal solution. 

\begin{theorem}\label{thm:existence}
    Consider the controlled adoption-opinion model \eqref{eq:compact_system}, with initial condition $y(0)=y_0$. 
    Then Problem~\ref{prob:cumulative_adoption} admits an optimal control $u_1^*\in\mc U$, and Problem~\ref{prob:terminal_adoption} admits an optimal control $u_2^*\in\mc U_{k_0}$. 
    Namely, there exist $u_1^*$ and $u_2^*$ such that 
     $$ J_1(u_1^*)=\max_{u\in\mc U}J_1(u), \quad J_2(u_2^*)= \! \! \max_{\substack{u\in\mc U_{k_0}\\ \scalebox{0.7}{$\sum$}_{t=0}^{T-1}u(t)^\top \! Q u(t)\le B}}\!\!\!\!\!\! J_2(u). $$
\end{theorem}
The proof can be found in Appendix~\ref{apx:proof_thm_existence}.

\subsection{Necessary optimality conditions}
We now derive necessary optimality conditions for Problems~\ref{prob:cumulative_adoption} and~\ref{prob:terminal_adoption} by applying the discrete-time version of the Pontryagin maximum principle \citep{bertsekas2012dynamic}. For both problems, we introduce a costate sequence $\mu(t)$ having the same dimension and ordering as the state $y(t)$, namely
$ \mu(t)=\left(\mu_s(t),\{\mu_{a^{[k]}}(t)\}_{k},\{\mu_{d^{[k]}}(t)\}_{k}, \{\mu_{x^{[k]}}(t)\}_{k}\right)$. 
Thus, $\mu_{x_i^{[k]}}(t)$ denotes the component of the costate associated with the opinion variable $x_i^{[k]}(t)$.
Since the control $u_i^{[k]}(t)$ enters the dynamics exclusively through the opinion equation \eqref{eq:dynamic_controlled_x}, the partial derivative of $x_i^{[k]}(t+1)$ with respect to the control is
\be \frac{\partial x_i^{[k]}(t+1)}{\partial u_i^{[k]}(t)} = (1-\lambda_i^{[k]}-\xi_i^{[k]})(1-p_i^{[k]}), \label{eq:dx_du}     \ee
which is nonnegative for all $i\in\mc V$ and $k\in\Kset$. Moreover, the quadratic cost term satisfies
 \be \frac{\partial}{\partial u_i^{[k]}}\!\left(-u(t)^\top Q\, u(t)\right) = -2q_i^{[k]}u_i^{[k]}(t).\label{eq:dQu_du}     \ee

\textbf{Problem~\ref{prob:cumulative_adoption}} The Hamiltonian is 
$\mc H_{1,t}(y(t),u(t),\mu(t+1\!)\!) := \ell_1(y(t),u(t)) \!+\! \mu(t+1)^\top \!\!\! f(y(t),u(t))$.
If $u_1^*$ is optimal with the corresponding optimal trajectory $y^*$, then there exists a costate sequence $\mu(t)$ satisfying the terminal condition $\mu(T)=\boldsymbol 0$,
and the adjoint equation $\mu(t)= \nabla_y\mc H_{1,t}(y^*(t),u_1^*(t),\mu(t+1))$, for $t=T-1,\ldots,0.$ 
Moreover, the optimal control satisfies
\beu u_1^*(t) \in \arg\max_{u\in U} \mc H_{1,t}(y^*(t),u,\mu(t+1)),\quad t=0,\dots,T-1.\label{eq:max_condition_cumulative}  \eeu
Differentiating $\mc H_{1,t}$ with respect to $u_i^{[k]}(t)$ and using \eqref{eq:dx_du}-\eqref{eq:dQu_du} gives
\beu \frac{\partial \mc H_{1,t}}{\partial u_i^{[k]}} \!=\! -2q_i^{[k]}u_i^{[k]}(t) + (1-\lambda_i^{[k]}-\xi_i^{[k]})(1-p_i^{[k]}) \mu_{x_i^{[k]}}(t+1). \eeu
For a fixed state-costate pair, maximizing the Hamiltonian with respect to $u_i^{[k]}(t)$ gives the unconstrained pointwise maximizer
\be \label{eq:uhat_P1}\widehat u_{1,i}^{[k]}(t) = \frac{1} {2q_i^{[k]}} (1-\lambda_i^{[k]}-\xi_i^{[k]})(1-p_i^{[k]})\mu_{x_i^{[k]}}(t+1).\ee
Hence, the pointwise maximizer entering the PMP conditions is
\be u_{1,i}^{[k]*}(t)=\min\Big\{1,\max\big\{0,\widehat u_{1,i}^{[k]}(t)\big\}\Big\},\quad i\in\mc V,\ k\in\Kset.\label{eq:optimal_control_P1} \ee
For a fixed state-costate pair, $\mc H_{1,t}$ is strictly
concave in $u(t)$ because of the quadratic term $-u(t)^\top Q u(t)$. Therefore, the maximizer \eqref{eq:optimal_control_P1} is unique. This does not imply global optimality of the resulting control sequence for the nonlinear dynamic optimization problem.

\textbf{Problem~\ref{prob:terminal_adoption}} 
Since the horizon is finite, Problem~\ref{prob:terminal_adoption} can be reduced to a
finite-dimensional nonlinear program in the control sequence. For the budget constraint $g(u):=\sum_{t=0}^{T-1}u(t)^\top Q u(t)-B\le0$, if the constraint is active and $B>0$, then $u\neq0$ and $\nabla g(u)\neq0$. Together with the box constraints, this ensures a standard constraint qualification, so the necessary conditions can be written in normal form, with the multiplier of the terminal payoff normalized to one.
The necessary optimality conditions involve a multiplier $\eta\ge 0$ associated with the budget constraint \eqref{eq:budget-constraint}, then the augmented Hamiltonian is $\mc H_{2,t}(y(t),u(t),\mu(t+1)) :=\! \mu(t+1)^\top \! f(y(t),u(t)) - \eta u(t)^\top \! Q u(t)$.
The terminal condition is given by $\mu_{a_i^{[k_0]}}(T)=1$, for all $i \in \mc V$,
while all the other terminal costate components are equal to zero. The adjoint equation is
$\mu(t)=\nabla_y\mc H_{2,t}(y^*(t),u_2^*(t),\mu(t+1))$, for all $t=T-1,\dots,0$. 
The optimal control satisfies the pointwise maximum condition
\beu  u_2^*(t) \in \arg\max_{u\in U_{k_0}} \mc H_{2,t}(y^*(t),u,\mu(t+1)), \quad t=0,\dots,T-1, \label{eq:max_condition_terminal_budget} \eeu
together with the complementarity conditions
\be\begin{aligned} \eta & \ge 0, \quad  \sum\nolimits_{t=0}^{T-1}u_2^*(t)^\top Q u_2^*(t)\le B, \\
&\eta\Big( \sum\nolimits_{t=0}^{T-1}u_2^*(t)^\top Q u_2^*(t)-B \Big)=0. \label{eq:budget_complementarity}\end{aligned} \ee
Differentiating $\mc H_{2,t}$ with respect to $u_i^{[k_0]}(t)$ and using \eqref{eq:dx_du}-\eqref{eq:dQu_du} gives
\begin{multline*}
    \frac{\partial \mc H_{2,t}}{\partial u_i^{[k_0]}}= -2\eta q_i^{[k_0]}u_i^{[k_0]}(t)\\[-2ex]
    +(1-\lambda_i^{[k_0]}-\xi_i^{[k_0]})(1-p_i^{[k_0]}) \mu_{x_i^{[k_0]}}(t+1).
\end{multline*}
Since the control is restricted to the target technology, the maximization is performed only over the components $u_i^{[k_0]}(t)$. When $\eta>0$, the unconstrained maximizer is
$$ \widehat u_{2,i}^{[k_0]}(t)=\frac{1}{2\eta q_i^{[k_0]}} (1-\lambda_i^{[k_0]}-\xi_i^{[k_0]})(1-p_i^{[k_0]})\mu_{x_i^{[k_0]}}(t+1),$$
and the corresponding pointwise maximizer is
\be u_{2,i}^{[k_0]*}(t)=\min\Big\{1,\max\big\{0,\widehat u_{2,i}^{[k_0]}(t)\big\}\Big\},\qquad i\in\mc V,\label{eq:optimal_control_P2} \ee
while $u_{2,i}^{[k]*}(t)=0$, for all $i\in\mc V$ and $k\neq k_0$. 
For a PMP extremal, the multiplier $\eta$ must satisfy the complementarity conditions \eqref{eq:budget_complementarity}. If the budget constraint is active, then $\sum_{t=0}^{T-1}u_2^*(t)^\top Q u_2^*(t)=B$. If the budget constraint is inactive, complementarity implies $\eta=0$. More generally, when $\eta=0$, the Hamiltonian is affine in the control and the pointwise maximizer need not be unique.
It is worth noting that the adoption variables do not appear explicitly in the feedback expressions \eqref{eq:optimal_control_P1} and \eqref{eq:optimal_control_P2}. This is because the control does not act directly on adoption, but only on the controlled baseline opinions. The effect of adoption on the computed intervention is encoded in the costate variables. In particular, the component $\mu_{x_i^{[k]}}(t+1)$ represents the marginal value, with respect to the remaining objective, of increasing the opinion of agent $i$ toward technology $k$ at the next time step. 


\subsection{Numerical computation of extremals}
\begin{figure}[t]
  \centering
  \includegraphics[width=0.95\linewidth]{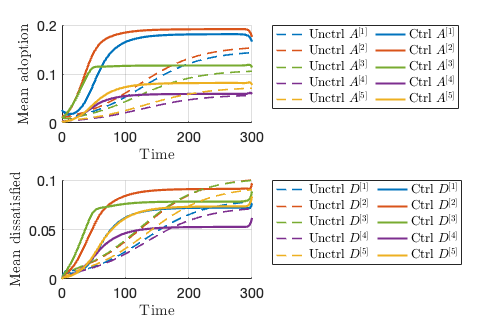}
  \caption{Problem~\ref{prob:cumulative_adoption} ($m=5$, $T=300$): mean adoption (top) and mean dissatisfied fraction (bottom) for each technology, uncontrolled (dashed) vs.\ controlled (solid).}
  \label{fig:prob1}
\end{figure}

\begin{figure}[t]
  \centering
  \includegraphics[width=0.95\linewidth]{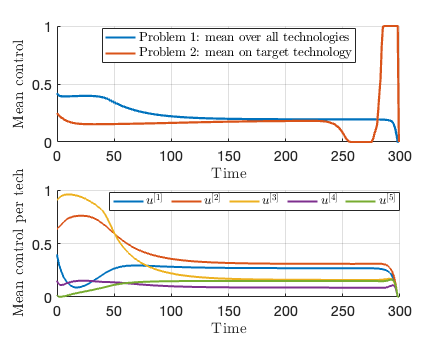}
  \caption{Average control effort for Problems 1 and 2 (top), and per-alternative allocation for Problem 1 (bottom).
  }
  \label{fig:prob1_control}
\end{figure}

\begin{figure}[t]
  \centering
  \includegraphics[width=0.8\linewidth]{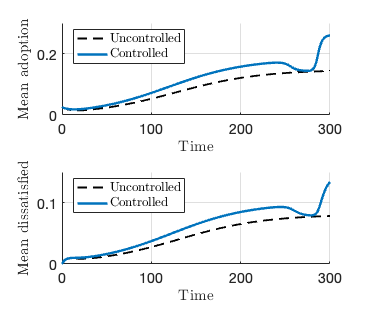}
  \caption{Problem~\ref{prob:terminal_adoption} with target $k_0=1$, $T=300$: mean adoption (top) and mean dissatisfied fraction (bottom), uncontrolled (dashed) vs.\ controlled (solid).}
  \label{fig:prob23}
\end{figure}

The necessary optimality conditions derived above define a coupled forward-backward system: the state dynamics is initialized at $y(0)=y_0$, whereas the costate dynamics is solved backward from the corresponding terminal condition. We compute numerical extremals satisfying these conditions using a forward-backward sweep method \citep{mcasey2012convergence}. Starting from an admissible control sequence $u^{(0)}$, at each iteration $h$ we first propagate the state dynamics forward in time and then the costate dynamics backward. The control is updated using the projected Hamiltonian maximizer (\eqref{eq:optimal_control_P1} for Problem~\ref{prob:cumulative_adoption}, \eqref{eq:optimal_control_P2} for Problem~\ref{prob:terminal_adoption}).
To improve numerical stability, we use the relaxed update 
\be u^{(h+1)}(t)=\omega \widetilde u^{(h+1)}(t)+(1-\omega)u^{(h)}(t),\quad \omega\in(0,1], \label{eq:relaxed_update} \ee
where $\widetilde u^{(h+1)}(t)$ is obtained from the corresponding projected maximizer using the current state and costate trajectories. 
Iterations are repeated until the control variation falls below a prescribed tolerance. The forward--backward sweep is used here as a numerical method to solve the PMP necessary conditions; no convergence or global-optimality guarantee is claimed, and the computed extremal may depend on the initialization.
For Problem~\ref{prob:terminal_adoption}, when the budget constraint is active, the multiplier $\eta>0$ is determined through an outer scalar search enforcing $\sum_{t=0}^{T-1}u(t)^\top Qu(t)=B$.\\
We now illustrate the two optimal control problems on a synthetic network of $n=10$ communities with $m=5$ competing alternatives. The interaction matrices and model parameters are drawn randomly in the admissible ranges and condition~\eqref{eq:hyp-unique} is verified numerically. The time horizon is $T=300$; the cost matrix $Q$ is diagonal with $q_i^{[k]}=0.8$ for all $i,k$. An approximate PMP extremal is computed via the forward-backward sweep algorithm~\eqref{eq:relaxed_update}. 
Figure~\ref{fig:prob1} shows, for each technology $k$, the mean adoption $\frac{1}{n}\sum_i a_i^{[k]}(t)$ and mean dissatisfied fraction $\frac{1}{n}\sum_i d_i^{[k]}(t)$ under uncontrolled (dashed) and controlled (solid) dynamics for the Problem~\ref{prob:cumulative_adoption}. In this simulation, the computed intervention increases the adoption level of every technology simultaneously, while generally reducing the dissatisfied fractions in the reported simulation. In this simulation, alternatives with lower dissatisfaction rates receive stronger intervention and exhibit larger adoption gains, consistently with the corresponding costate profiles.
Figure~\ref{fig:prob1_control} displays the average control effort over time and its allocation across alternatives. The controller initially concentrates resources on the most responsive alternatives, then progressively rebalances as adoption levels saturate. This time-varying reallocation reflects the backward propagation of the costate: as adoption of a technology approaches its controlled equilibrium, the marginal value $\mu_{x_i^{[k]}}$ decreases and the computed control effort shifts toward less saturated alternatives.
Figure~\ref{fig:prob23} shows the mean adoption and dissatisfied fraction for the target technology $k_0=1$ under Problem~\ref{prob:terminal_adoption}. In this case, the intervention acts exclusively on the opinion toward the target technology and the available budget is allocated nonuniformly over the horizon. In particular, the control becomes more concentrated near the terminal time. This behavior is consistent with the terminal objective: interventions applied too early may increase adoption initially but also leave more time for adopters to become dissatisfied and subsequently switch to competing technologies before time $T$. Concentrating intervention closer to the end of the horizon limits these losses while still accounting for the two-step delay between control and adoption described in Remark \ref{rem:control}. The controlled dissatisfied fraction associated with $k_0$ may consequently exceed its uncontrolled counterpart. Indeed, the increase in $a^{[k_0]}$ generates a larger inflow $\delta^{[k_0]}a^{[k_0]}$ into $D^{[k_0]}$, whereas its outflow is driven by opinions toward the competing technologies and is not directly increased by the target-specific intervention.
\paragraph*{Code availability.}
All numerical experiments can be reproduced using the code and parameter sets available at \url{https://github.com/martinalutto/competing-technologies-simulations}.

\section{Conclusion}\label{sec:conclusion}
We proposed a networked adoption-opinion framework for multiple competing alternatives, jointly capturing social influence, adoption, dissatisfaction, and opinion-dependent switching. We established existence of an adoption-diffused equilibrium and provided sufficient conditions for uniqueness and convergence. The comparison between $m=2$ and $m>2$ showed that, with multiple alternatives, equilibrium adoption generally depends on the broader competitive environment. We also formulated finite-horizon opinion-shaping control problems, derived necessary optimality conditions, and computed numerical PMP extremals.\\
Two main limitations remain. First, the model has not yet been identified or validated against empirical data, so the results are structural rather than predictive. Second, the current compartmental structure does not capture permanent exit or true extinction of an alternative. Future work will address these aspects, less conservative uniqueness and convergence conditions, and time-varying interaction networks.

\appendix

\section{Proofs of the results in Section~\ref{sec:model}}\label{apx:proof_sec_2}
\subsection{Proof of Proposition \ref{prop:invariant_general}}
Assume that $y(t) \in \X$ at a time instant $t \geq0$. We show that also $y(t+1)$ belongs to $\X$.
We first prove non-negativity of the adoption variables. For every $i\in\V$, from~\eqref{eq:gen_s}, we have
$$s_i(t+1)= s_i(t)\Big(1-\! \sum\nolimits_{k \in \Kset}\beta_i^{[k]}x_i^{[k]}(t) \! \sum\nolimits_{j\in \mc V} \!\! W_{ij}a_j^{[k]}(t)\! \Big)\ge 0. $$
For every $i\in\V$ and $k\in\Kset$, considering ~\eqref{eq:gen_a} we have $a_i^{[k]}(t+1)
\geq (1-\delta_i^{[k]})a_i^{[k]}(t)$, and from $\delta_i^{[k]}\in[0,1]$, $a_i^{[k]}(t+1)\geq 0$. Similarly, from \eqref{eq:gen_d}, using $x_i^{[\ell]}(t)\in[0,1]$ and Assumption~\ref{ass:main}(iii), we get $d_i^{[k]}(t+1)\ge 0$.
Summing \eqref{eq:gen_s}--\eqref{eq:gen_d} over $k\in\Kset$ yields $s_i(t+1)+\sum\nolimits_{k\in\Kset}\big(a_i^{[k]}(t+1)+d_i^{[k]}(t+1)\big)
=s_i(t)+\sum\nolimits_{k\in\Kset}\big(a_i^{[k]}(t)+d_i^{[k]}(t)\big)=1$,
for all $i \in \mc V$.
Finally, from \eqref{eq:gen_x} and row-stochasticity of $\widetilde{W}$ and $W$, one has $x^{[k]}(t+1)\in[0,1]^n$ when $x^{[k]}(t)\in[0,1]^n$ and $a^{[k]}(t)\in[0,1]^n$. We conclude that $y(t+1)$ belongs to $\X$. By induction, the trajectory remains in $\X$ for all $t\geq 0$.

\subsection{Proof of Corollary \ref{cor:monotone_s}}
(i) By Proposition \ref{prop:invariant_general}, the subtracted term in \eqref{eq:gen_s} is nonnegative for all $i\in\V$. Hence $s_i(t+1)\leq s_i(t)$ for all $i\in\V$, which proves the claim.
(ii) From the opinion dynamics in \eqref{eq:gen_x}, we get $ x^{[k]}(t+1) \geq (I-\Lambda^{[k]}-\Xi^{[k]})p^{[k]} > \boldsymbol{0}$ for all $t\geq0$ and $k \in \Kset$.


\section{Proofs of the results in Section~\ref{sec:equilibria_general}}\label{apx:proof_sec_3}
\subsection{Proof of Proposition \ref{prop:afe}}
With $a^{[k]}=\boldsymbol{0}$ and $d^{[k]}=\boldsymbol{0}$, the adoption dynamics are trivially satisfied. Equation \eqref{eq:gen_x} reduces to the linear fixed point in equation \eqref{eq:x_afe}. 
Since $\widetilde W$ is row-stochastic and $\lambda_i^{[k]}+\xi_i^{[k]}<1$ for every $i$,
$\rho(\Lambda^{[k]}\widetilde W) \leq \|\Lambda^{[k]}\widetilde W\|_\infty = \max_i\lambda_i^{[k]}<1$.
Hence, $I-\Lambda^{[k]}\widetilde W$ is invertible.
Moreover, $x_{\rm AF}^{[k]}>\boldsymbol{0}$ since $(I-\Lambda^{[k]}\widetilde W)^{-1}\ge0$ and $(I-\Lambda^{[k]}-\Xi^{[k]})p^{[k]}>\boldsymbol{0}$.
Similarly, since $p^{[k]}\leq\boldsymbol 1$,
\begin{align}
    (I-\Lambda^{[k]}-\Xi^{[k]})p^{[k]}&\leq (I-\Lambda^{[k]}-\Xi^{[k]})\boldsymbol 1 \label{eq:relations}\\
    &\leq (I-\Lambda^{[k]})\boldsymbol 1=(I-\Lambda^{[k]}\widetilde W)\boldsymbol 1,\nonumber    
\end{align}
where the last equality follows from $\widetilde W\boldsymbol 1=\boldsymbol 1$. Since $(I-\Lambda^{[k]}\widetilde W)^{-1}$ is nonnegative, we obtain $x_{AF}^{[k]}\leq\boldsymbol 1$.\\
Now fix any $\varepsilon\in(0,1)$ and choose an initial condition arbitrarily close to $y_{AF}$ by setting $s(0)=(1-\varepsilon)\boldsymbol{1}$, $a^{[k_0]}(0)=\varepsilon\boldsymbol{1}$ for some $k_0\in\Kset$, with all other adoption and dissatisfaction variables equal to $0$, and let $x^{[k]}(0)=x_{AF}^{[k]}$ for all $k\in\Kset$. By Corollary~\ref{cor:monotone_s}, $s(t)\le s(0)$ for all $t$, hence $\|s(t)-\boldsymbol{1}\|_\infty\ge \varepsilon$ for all $t \geq 0$. It follows that 
$ \|y(t)-y_{AF}\|_\infty \ \ge\ \|s(t)-\mathbf 1\|_\infty \ \ge\ \varepsilon\quad \forall t\ge 0. $
Since $\varepsilon>0$ can be chosen arbitrarily small, these initial conditions can be made arbitrarily close to $y_{AF}$, while the corresponding trajectories do not converge to $y_{AF}$. Thus, $y_{AF}$ is not attractive and, consequently, it cannot be asymptotically stable.

\subsection{Proof of Theorem \ref{thm:exist_general}}
The proof proceeds in three steps. First, we identify a compact set containing the equilibrium opinion profiles. Second, for any fixed opinion profile, we uniquely determine the corresponding adoption and dissatisfaction variables. Finally, we formulate the equilibrium conditions as a continuous map and apply Brouwer's fixed-point theorem.

At equilibrium, \eqref{eq:gen_x} yields \eqref{eq:x_star_general}. Since $(I-\Lambda^{[k]}\widetilde W)^{-1}$ is well defined and nonnegative and $a^{[k]\star}\ge0$, we have $x^{[k]\star}\ge x_{AF}^{[k]}>0$.
Moreover, any admissible equilibrium satisfies $x^{[k]\star}\le\boldsymbol{1}$. Then, $x^*$ necessarily belongs to the nonempty, compact, and convex set
$\mathcal C:=\prod_{k\in\Kset}\big[x_{AF}^{[k]},\boldsymbol{1}\big]=\left\{x\in[0,1]^{nm}:x_{AF}^{[k]}\le x^{[k]}\le \boldsymbol{1}\right\}.$
We next show that, once $x^\star\in\mathcal C$ is assigned, the corresponding equilibrium values of $a^\star$ and $d^\star$ are uniquely determined.  
Substituting $s_i^\star=0$ into \eqref{eq:gen_a}--\eqref{eq:gen_d} at equilibrium, we obtain
\begin{subequations}
\begin{align}
    \delta_i^{[k]}a_i^{[k]\star} &= \gamma_i^{[k]}x_i^{[k]\star}\mspace{-10mu} \sum_{\ell\in\Kset\setminus\{k\}} \mspace{-15mu} d_i^{[\ell]\star}, \label{eq:eq_ad}\\
    \delta_i^{[k]}a_i^{[k]\star} &= \Bigl(\!\sum_{\ell\in\Kset\setminus\{k\}}\mspace{-15mu} \gamma_i^{[\ell]}x_i^{[\ell]\star}\Bigr)d_i^{[k]\star}, \label{eq:eq_da}  
\end{align}\end{subequations}
for every $i\in\V$ and $k\in\Kset$.
From \eqref{eq:def-bar-dx}, 
equating \eqref{eq:eq_ad} and \eqref{eq:eq_da} gives $\gamma_i^{[k]}x_i^{[k]\star}\bigl(\bar d_i^\star-d_i^{[k]\star}\bigr)=\bigl(\bar x_i^\star-\gamma_i^{[k]}x_i^{[k]\star}\bigr)d_i^{[k]\star}$, 
hence
\vspace{-2pt}\begin{equation}\label{eq:d_share_proof}
    d_i^{[k]\star}=\frac{\gamma_i^{[k]}x_i^{[k]\star}}{\bar x_i^\star}\bar d_i^\star.
\end{equation}
Substituting \eqref{eq:d_share_proof} into \eqref{eq:eq_da}, we obtain
\begin{equation}\label{eq:a_formula_proof}
    a_i^{[k]\star} = \frac{\gamma_i^{[k]}x_i^{[k]\star}}{\delta_i^{[k]}} \Big(1-\frac{\gamma_i^{[k]}x_i^{[k]\star}}{\bar x_i^\star}\Big)\bar d_i^\star .
\end{equation}
Since $s_i^\star=0$, the normalization constraint in $\X$ at equilibrium becomes $\sum\nolimits_{k\in\Kset}\bigl(a_i^{[k]\star}+d_i^{[k]\star}\bigr)=1$, for all $i\in\V.$
Using \eqref{eq:d_share_proof} and \eqref{eq:a_formula_proof}, for every $i\in\V$ this yields
\beu \bar d_i^\star \sum\nolimits_{k\in\Kset} \Big( \frac{\gamma_i^{[k]}x_i^{[k]\star}}{\delta_i^{[k]}}\Big(1-\frac{\gamma_i^{[k]}x_i^{[k]\star}}{\bar x_i^\star}\Big) +\frac{\gamma_i^{[k]}x_i^{[k]\star}}{\bar x_i^\star}\Big)=1,  \eeu
that is, using \eqref{eq:def-bar-dx}, 
\be \label{eq:D_formula_proof}
\bar d_i^\star = \Bigg(\sum\nolimits_{k\in\Kset} \frac{\gamma_i^{[k]}x_i^{[k]\star}}{\delta_i^{[k]}}\! \Big(1-\frac{\gamma_i^{[k]}x_i^{[k]\star}}{\bar x_i^\star}\Big) +1\Bigg)^{-1}\!\!\!. \ee
Thus, for every $x\in\mathcal C$, the vectors $a(x)$ and $d(x)$ are uniquely determined, nonnegative, and continuous in $x$. By construction, they satisfy the normalization constraint, and hence $a^{[k]}(x),d^{[k]}(x)\in[0,1]^n$.
We can now define the map $L:\mathcal C\to[0,1]^{nm}$ blockwise as
\begin{equation}\label{eq:L_general}
    \! L^{[k]}\mspace{-3mu}(x):= (I-\Lambda^{[k]}\!\widetilde{W})^{-1}\!(\!(I-\Lambda^{[k]}\mspace{-3mu}-\Xi^{[k]}\!)p^{[k]}\!+\Xi^{[k]}\mspace{-3mu}W\mspace{-3mu}a^{[k]}\mspace{-3mu} (x)\!).
\end{equation}
By continuity of $a^{[k]}(x)$, the map $L$ is continuous on $\mathcal C$. We now show that $L(\mc C)\subseteq \mc C$. For every $x\in\mathcal C$, relations \eqref{eq:a_formula_proof} and \eqref{eq:D_formula_proof} give $a^{[k]}(x)\ge \boldsymbol 0$. Since $a^{[k]}(x)\le\boldsymbol{1}$ and $W$ is row-stochastic, $Wa^{[k]}(x)\le\boldsymbol{1}$. 
Together with \eqref{eq:relations}, we obtain $x_{AF}^{[k]}\le L^{[k]}(x)\le \boldsymbol 1$, for all $k\in\Kset$. Therefore, $L(x)\in\mc C$ for every $x\in\mc C$. Since $\mc C$ is nonempty, compact, and convex, Brouwer's fixed-point theorem guarantees the existence of $x^\star\in\mc C$ such that $L(x^\star)=x^\star$. Defining $a^\star:=a(x^\star)$ and $d^\star:=d(x^\star)$ through \eqref{eq:d_share_proof}--\eqref{eq:D_formula_proof}, relations \eqref{eq:eq_ad} and \eqref{eq:eq_da} are satisfied. Hence, $y^\star$ is an equilibrium of the system.
Finally, since $x^\star\in\mc C$, we have $x^{[k]\star}>\boldsymbol 0$ for all $k$, hence $\bar x_i^\star>0$.
Moreover, $\bar x_i^\star-\gamma_i^{[k]}x_i^{[k]\star} =\sum\nolimits_{\ell\in\Kset\setminus\{k\}} \gamma_i^{[\ell]}x_i^{[\ell]\star}>0.$
By \eqref{eq:d_share_proof} and \eqref{eq:a_formula_proof}, we get $d_i^{[k]\star}>0$, $a_i^{[k]\star}>0$ for all $i\in\V$, $k\in\Kset$. This concludes the proof.
\subsection{Proof of Proposition \ref{prop:no_partial_general}}
Let $\mc J := \{i \in \mathcal V: s_i^\star=0\}$. By assumption, $\mc J \neq \emptyset$. Suppose by contradiction $\mc J \neq \mc V$. For any $i\in\mathcal J$, since $s_i^\star=0$, we have $\sum_{k\in\mathcal K}(a_i^{[k]\star}+d_i^{[k]\star})=1$. 
At any equilibrium, \eqref{eq:gen_x} and Assumption~\ref{ass:main}(ii) give $x_i^{[k]\star}>0$ for all $i\in\V$ and $k\in\Kset$. Since $\gamma_i^{[k]}>0$, it follows that $\gamma_i^{[k]}x_i^{[k]\star}>0$.
If $a_i^{[k]\star}=0$ for all $k$, then from \eqref{eq:gen_a} at equilibrium, we obtain $\sum_{\ell\in\Kset\setminus\{k\}}d_i^{[\ell]\star}=0$ for all $k$. This implies $d_i^{[k]\star}=0$ for all $k$, contradicting the normalization constraint. Hence, there exists $k_0\in\mathcal K$ such that $a_i^{[k_0]\star}>0$.
Since $W$ is irreducible and $\mc J$ is a nonempty proper subset of $\mc V$, there exist $p\in\mc J$ and $q \notin\mc J$ such that $W_{qp}>0$. As shown above, there exists $k_0\in\Kset$
such that $a_p^{[k_0]\star}>0$. Moreover, since $q\notin J$, we have $s_q^\star>0$. Evaluating \eqref{eq:gen_s} at equilibrium gives
$$0=s_q^\star\sum\nolimits_{k\in\mc K}\beta_q^{[k]}x_q^{[k]\star}\sum\nolimits_{r\in\mc V}W_{qr}a_r^{[k]\star}.$$
However, the term corresponding to $k=k_0$ satisfies $s_q^\star\beta_q^{[k_0]}x_q^{[k_0]\star}W_{qp}a_p^{[k_0]\star}>0$, that is a contradiction. Therefore, $\mc J=\mc V$, and hence $s^\star=0$.
\subsection{Proof of Corollary \ref{cor:equilibrium_dichotomy}}
Let $y^\star$ be an equilibrium. If $s_i^\star=0$ for some $i\in\V$, Proposition~\ref{prop:no_partial_general} gives $s^\star=\boldsymbol{0}$. Since
$x^{[k]\star}>\boldsymbol{0}$ and $\delta_i^{[k]}>0$, the equilibrium relations \eqref{eq:eq_ad}-\eqref{eq:a_formula_proof}, together with the
normalization constraint, imply $a^{[k]\star},d^{[k]\star}>\boldsymbol{0}$ for every $k$. Hence $y^\star$ is an ADE.
Otherwise, $s_i^\star>0$ for every $i$. Equation \eqref{eq:gen_s} and
the positivity of $\beta_i^{[k]}x_i^{[k]\star}$ imply $Wa^{[k]\star}=\boldsymbol{0}$ for every $k$. Since $W$ is irreducible and $a^{[k]\star}\ge\boldsymbol{0}$, it follows that $a^{[k]\star}=\boldsymbol{0}$. Equation \eqref{eq:gen_a} then gives $d^{[k]\star}=\boldsymbol{0}$ for every $k$, and normalization yields $s^\star=\boldsymbol{1}$. Finally, \eqref{eq:gen_x} gives
$x^{[k]\star}=x_{\rm AF}^{[k]}$, hence $y^\star=y_{\rm AF}$.

\subsection{Proof of Theorem \ref{thm:unique}}
We establish uniqueness of the ADE by showing that the equilibrium map $L$, defined in \eqref{eq:L_general}, is a contraction. 

We endow $\mathcal C\subset\mathbb{R}^{nm}$ with the product
$\ell_\infty$-norm $\|x\|_\infty:=\max_{k\in\Kset}\|x^{[k]}\|_\infty $.
For every $i\in\V$ and $k\in\Kset$, define
\beu g_i^{[k]}(x):=\frac{\gamma_i^{[k]}x_i^{[k]}}{\delta_i^{[k]}}\left(1-\frac{\gamma_i^{[k]}x_i^{[k]}}{\bar x_i}\right),\qquad \bar x_i:=\sum_{r\in\Kset}\gamma_i^{[r]}x_i^{[r]}. \vspace{-5pt} \eeu
Then, from \eqref{eq:D_formula_proof}--\eqref{eq:a_formula_proof},
$ \bar d_i(x)=\left(1+\sum\nolimits_{r\in\Kset}g_i^{[r]}(x)\right)^{-1}$, $a_i^{[k]}(x)=g_i^{[k]}(x)\bar d_i(x)$.
We first derive a Lipschitz bound for $a^{[k]}$. Since for each $i\in\mc V$, the component $a_i^{[k]}(x)$ depends only on the variables $(x_i^{[\ell]})_{\ell\in\Kset}$, it is enough to estimate the partial derivatives $\partial a_i^{[k]}(x) / \partial x_i^{[\ell]}$ for all $\ell\in\Kset$.
Direct differentiation gives
\begin{subequations}
\begin{align} \label{eq:der_num_k}
\frac{\partial g_i^{[k]}}{\partial x_i^{[k]}} 
&=\frac{\gamma_i^{[k]}}{\delta_i^{[k]}} \left( 1-\frac{\gamma_i^{[k]}x_i^{[k]}}{\bar x_i}\right)^2,\quad \ell=k,\\
\frac{\partial g_i^{[k]}}{\partial x_i^{[\ell]}} & =\frac{\gamma_i^{[\ell]}}{\delta_i^{[k]}}\left(\frac{\gamma_i^{[k]}x_i^{[k]}}{\bar x_i}\right)^2,\quad \ell\neq k.
\end{align}
\end{subequations}
Since $0\leq\gamma_i^{[r]}x_i^{[r]}/\bar x_i\leq1$, it follows that
 \be \label{eq:bound_der_num_general}
\left| \frac{\partial g_i^{[k]}}{\partial x_i^{[\ell]}} 
\right| \le \frac{\gamma_i^{[\ell]}}{\delta_i^{[k]}}, \ee
for every $k,\ell\in\Kset$. Moreover, differentiating $\bar d_i$ yields
 \be \label{eq:der_den}\begin{aligned}
    \frac{\partial}{\partial x_i^{[\ell]}} \!\Big(\!1+\!\sum_{r\in\Kset} g\big(x_i^{[r]}\big)\!\Big) = \frac{\gamma_i^{[\ell]}}{\delta_i^{[\ell]}} - \frac{2(\gamma_i^{[\ell]}\!)^2x_i^{[\ell]}} {\delta_i^{[\ell]}\bar x_i} + \frac{\gamma_i^{[\ell]}}{\left(\bar x_i \right)^2}\!\sum_{r\in\Kset}\frac{(\gamma_i^{[r]}\! x_i^{[r]})^2\! }{\delta_i^{[r]}}. 
\end{aligned}\ee
From $\gamma_i^{[r]}x_i^{[r]}\leq \bar x_i$, one has
$$\left| \frac{\gamma_i^{[\ell]}}{\delta_i^{[\ell]}} - \frac{2(\gamma_i^{[\ell]})^2x_i^{[\ell]}}{\delta_i^{[\ell]}\bar x_i} \right| =  \frac{\gamma_i^{[\ell]}}{\delta_i^{[\ell]}} \left| 1 - \frac{2 \gamma_i^{[\ell]}x_i^{[\ell]}}{\bar x_i} \right| \le \frac{\gamma_i^{[\ell]}}{\delta_{i,\min}},$$
using the first definition in \eqref{eq:params-def}.
Moreover, regarding the third factor in \eqref{eq:der_den},
\begin{align}
    \mspace{-18mu} \frac{\gamma_i^{[\ell]}}{\left(\bar x_i\!\right)^2} \mspace{-3mu} \sum_{r\in\Kset}\! \frac{(\gamma_i^{[r]}x_i^{[r]})^2}{\delta_i^{[r]}} &\le\frac{\gamma_i^{[\ell]}}{\ds \delta_{i,\min}}\! \sum_{r\in\Kset}\! \Big(\frac{\gamma_i^{[r]}x_i^{[r]}}{\bar x_i}\!\Big)^2 \mspace{-5mu} \le\frac{\gamma_i^{[\ell]}}{\ds \delta_{i,\min}}. \label{eq:bound_square_term}
\end{align}
Hence, from \eqref{eq:der_den},
\begin{align}
\left|\frac{\partial}{\partial x_i^{[\ell]}} \Big(1+\sum\nolimits_{r\in\Kset} g\big(x_i^{[r]}\big)\Big)\right| \le \frac{2 \gamma_i^{[\ell]}}{\ds \delta_{i,\min}}.
\label{eq:bound_der_den}
\end{align}
Observe also that since, for each $r\in\Kset$, $(\gamma_i^{[r]}x_i^{[r]})^2 \le (\gamma_i^{[r]}x_i^{[r]}) \sum\nolimits_{h\in\Kset}\gamma_i^{[h]}x_i^{[h]}$,
it follows that
$$\frac{1}{\bar x_i} \sum\nolimits_{r\in\Kset}\frac{(\gamma_i^{[r]}x_i^{[r]})^2}{\delta_i^{[r]}} \le \sum\nolimits_{r\in\Kset}\frac{\gamma_i^{[r]}x_i^{[r]}}{\delta_i^{[r]}}.$$
Therefore,
\begin{align}
    1+\sum\nolimits_{r\in\Kset}\frac{\gamma_i^{[r]}x_i^{[r]}}{\delta_i^{[r]}} - \frac{1}{\bar x_i} \sum\nolimits_{r\in\Kset}\frac{(\gamma_i^{[r]}x_i^{[r]})^2}{\delta_i^{[r]}} \ge 1. \label{eq:den_lower_bound}
\end{align}
Then $\bar d_i \leq 1$ and by \eqref{eq:bound_der_den}, its partial derivatives will be bounded as
 \be \label{eq:partial-d}
    \mspace{-15mu} \left| \frac{\partial \bar d_i}{\partial x_i^{[\ell]}} \right| = \left| \frac{\frac{\partial}{\partial x_i^{[\ell]}} \Big(1+\sum_{r\in\Kset} g\big(x_i^{[r]}\big)\Big)}{\Big(1+\sum_{r\in\Kset} g\big(x_i^{[r]}\big)\Big)^2}\right| \leq \frac{2\gamma_i^{[\ell]}}{\ds \delta_{i,\min}}. \ee
While from $0\leq\gamma_i^{[r]}x_i^{[r]}/\bar x_i\leq1$ and the fact that $\theta(1-\theta)\le \frac14$, for all $\theta\in[0,1]$, we obtain
 \begin{align}
    0\le g_i^{[k]}(x) \le \frac{\bar x_i}{4\delta_i^{[k]}}\le \frac{1}{4\delta_i^{[k]}},
    \label{eq:num_upper_bound}  
\end{align}
where the last inequality follows from Assumption~\ref{ass:main}(iii).
By \eqref{eq:bound_der_num_general}, \eqref{eq:den_lower_bound}, \eqref{eq:partial-d} and \eqref{eq:num_upper_bound}, for all $\ell\in\Kset$ we get
\be \label{eq:partial_a_bound_final}\begin{aligned}
    \allowdisplaybreaks
    \mspace{-10mu}\left| \frac{\partial a_i^{[k]}}{\partial x_i^{[\ell]}}(x) \right| 
    &\leq \left| \frac{\partial g}{\partial x_i^{[\ell]}} \right| |\bar d_i|+\left| \frac{\partial \bar d_i}{\partial x_i^{[\ell]}}g\big(x_i^{[k]}\big) \right| \\
    &\le \frac{\gamma_i^{[\ell]}}{\delta_i^{[k]}} \!+\! \frac{1}{4\delta_i^{[k]}}\frac{2\gamma_i^{[\ell]}}{\ds \delta_{i,\min}} = \frac{\gamma_i^{[\ell]}}{\delta_i^{[k]}}\Big(\! 1\!+\! \frac{1}{2 \ds \delta_{i,\min}} \! \Big).    
\end{aligned}\ee
Since $a_i^{[k]}(x)$ depends only on $(x_i^{[\ell]})_{\ell\in\Kset}$, the mean-value theorem implies for all $x,y\in\mathcal C$,
$$\|a^{[k]}(x)-a^{[k]}(y)\|_\infty\le \max_{i\in\mc V} \sum\nolimits_{\ell\in\Kset} \sup_{z\in \mc C} \Big| \frac{\partial a_i^{[k]}}{\partial x_i^{[\ell]}}(z) \Big| \|x-y\|_{\infty}. $$
Using \eqref{eq:partial_a_bound_final} and Assumption \ref{ass:main}(iii), this yields
\vspace{-3pt} \begin{align}
    \mspace{-10mu}\|a^{[k]}\!(x)\!-\!a^{[k]}\!(y)\|_\infty \! & \ds \le \max_{i\in\mc V} \sum\nolimits_{\ell\in\Kset} \frac{\gamma_i^{[\ell]}}{\delta_i^{[k]}} \!\Big(\! 1\!+\!\frac{1}{2\ds \delta_{i,\min}}\! \Big) \|x \!-\! y\|_{\infty}\nonumber \\
    &\leq \max_{i\in\mc V}\frac{1}{\delta_i^{[k]}}\! \Big(1 \!+\! \frac{1}{2\ds \delta_{i,\min}}\Big) \! \|x\! -\! y\|_{\infty}. \label{eq:Lak}
\end{align}
The constant in \eqref{eq:Lak} is uniform over $x,y\in\mc C$.
We now use this estimate to show that the equilibrium opinion map $L$, defined in \eqref{eq:L_general}, is a contraction.
As already shown in the proof of Theorem~\ref{thm:exist_general}, one has $L(\mathcal C)\subseteq\mathcal C$.
For every $x,y\in\mathcal C$ and for each $k\in\Kset$,
$ L^{[k]}(x) - L^{[k]}(y) = (I-\Lambda^{[k]}\widetilde{W})^{-1}\Xi^{[k]}W (a^{[k]} (x)-a^{[k]} (y))$,
which in $\infty$-norm, using \eqref{eq:Lak}, yields
\beu \begin{aligned}
\|L^{[k]}(x)&-L^{[k]}(y)\|_\infty \\
&\leq \left\|(I-\Lambda^{[k]}\widetilde W)^{-1}\Xi^{[k]}W\right\|_\infty \|a^{[k]}(x)-a^{[k]}(y)\|_{\infty} \\
&\leq \frac{\max_i \xi_i^{[k]}}{1- \max_i \lambda_i^{[k]}} \max_{i\in\mc V}\frac{1}{\delta_i^{[k]}}\Big(\! 1\!+\! \frac{1}{2\ds \delta_{i,\min}}\Big) \! \|x-y\|_{\infty},
\vspace{-3pt}
\end{aligned} \eeu
since $\widetilde W$ is row-stochastic and $\lambda_i^{[k]}<1$ for all $i$.
Taking the maximum over $k\in\Kset$ and using \eqref{eq:hyp-unique}, there exists $c<1$ such that $\|L(x)-L(y)\|_\infty \leq c \|x-y\|_{\infty}$.
Hence, $L$ is a contraction on the complete metric space $\mathcal C$. By Banach's fixed-point theorem, $L$ admits a unique fixed point $x^\star\in\mathcal C$. Finally, by \eqref{eq:d_share_proof}-\eqref{eq:a_formula_proof}, once $x^\star$ is fixed, the vectors $d^\star$ and $a^\star$ are uniquely determined. Therefore, the corresponding equilibrium with $s^\star=\boldsymbol 0$ is unique.
\subsection{Proof of Theorem \ref{thm:convergence}}
We first observe that condition~\eqref{eq:hyp-convergence} implies the uniqueness condition~\eqref{eq:hyp-unique}.
Indeed, for every $k\in\Kset$,
\begin{align*}
        \max_i \frac{1}{\delta_{i}^{[k]}}\Big(\! 1\!+\! \frac{1}{\ds 2\delta_{i,\min}}\Big)& \leq \max_i \frac{1}{\delta_{i,\min}}\Big(\! 1\!+\! \frac{1}{\ds 2\delta_{i,\min}}\Big) \\[-1ex]
        &\leq\frac{3}{2 \min_i \delta_{i,\min}^2}<\frac{6}{\min_i \varepsilon_i}.
\end{align*}
where the second inequality uses $\delta_{i,\min}\le1$, while the third follows from
$\varepsilon_i =\underline{x}_i\delta_{i,\min}^2\gamma_{i,\min}^3 \le\delta_{i,\min}^2$. Multiplying by $\max_k\frac{\max_i\xi_i^{[k]}}
{1-\max_i\lambda_i^{[k]}}$ and using \eqref{eq:hyp-convergence} yield \eqref{eq:hyp-unique}.
Then, Theorem~\ref{thm:unique} guarantees that the ADE is unique.
It remains to prove convergence. We first establish that $y(t+1)-y(t)\to0$ and then show that every limit point coincides with the unique ADE.

By Proposition~\ref{prop:invariant_general}, the trajectory remains in the compact set $\X$. Moreover, by Corollary~\ref{cor:monotone_s}, $s(t)$ is non-increasing and bounded below by $\boldsymbol 0$, hence the limit $\bar s:=\lim_{t\to\infty}s(t)$ exists, with $\boldsymbol 0\le \bar s\le s(0)$. Hence, $s_i(t)-s_i(t+1)\to0$.
For each $i$, let $w_i(t):=\big[(a_i^{[k]}(t))_{k\in\Kset},(d_i^{[k]}(t))_{k\in\Kset}\big]\in\mathbb R^{2m}.$ Then \eqref{eq:gen_a}--\eqref{eq:gen_d} can be written as $w_i(t+1):=P_i(t)w_i(t)+b_i(t)$,  where
\vspace{-3pt}
$$P_i(t):=\begin{bmatrix}
\diag(1-\delta_i^{[k]})_k&T_i(t)\\[-0.5ex]
\diag(\delta_i^{[k]})_k&\diag\left(1-\sum_{\ell\neq k}\gamma_i^{[\ell]}x_i^{[\ell]}(t)\right)_k
\end{bmatrix},$$
with
\vspace{-2pt}$$[T_i(t)]_{k\ell}:=\begin{cases}
\gamma_i^{[k]}x_i^{[k]}(t),&k\neq\ell,\\
0,&k=\ell,
\end{cases}$$
and $b_i(t):=\left[\big(\beta_i^{[k]}x_i^{[k]}(t)s_i(t)(Wa^{[k]}(t))_i\big)_k,\mathbf0\right]$.
By Assumption~\ref{ass:main}, $P_i(t)$ is nonnegative and column-stochastic. Moreover, $\boldsymbol 1^\top b_i(t)=s_i(t)-s_i(t+1)\to0$.
Since $b_i(t)\geq\boldsymbol 0$, it follows that $b_i(t)\to\boldsymbol 0$, and therefore $b_i(t+1)-b_i(t)\to\boldsymbol 0$.
We now establish a uniform contraction property for products of three consecutive matrices $P_i(t)$. 
Given definitions in \eqref{eq:params-def}, from the opinion dynamics, $x_i^{[k]}(t)\geq\underline x_i$ for all $t\geq1$ and every $i,k$. Moreover, since $x_i^{[\ell]}(t)\leq1$ and $\sum_k\gamma_i^{[k]}\leq1$, it follows that for all $t\geq1$,
\begin{subequations}\label{eq:transition-lower-bounds}
    \begin{align}
        \gamma_i^{[k]}x_i^{[k]}(t)&\geq\gamma_{i,\min}\underline x_i,\\
        1-\sum_{\ell\neq r}\gamma_i^{[\ell]}x_i^{[\ell]}(t)\geq 1-&\sum_{\ell\neq r}\gamma_i^{[\ell]}\geq\gamma_i^{[r]}\geq\gamma_{i,\min}.    
    \end{align}
\end{subequations}  
Fix any $r\in\Kset$. From every state $A_k$ or $D_k$ there exists a three-step path to $D_r$. If $k=r$, use $A_r\to D_r\to D_r\to D_r$ and $D_r\to D_r\to D_r\to D_r$; if $k\neq r$, use $A_k\to D_k\to A_r\to D_r$ and $D_k\to A_r\to D_r\to D_r$. Each of these paths corresponds to a product of transition probabilities along the edges, and by Assumption~\ref{ass:main} together with the lower bounds in \eqref{eq:transition-lower-bounds}, every transition along such paths is bounded below by a positive time-independent variable.
In particular, their weights are bounded below by $\delta_{i,\min}\gamma_{i,\min}^2$, $\gamma_{i,\min}^3$, $\underline x_i\delta_{i,\min}^2\gamma_{i,\min}$, $\underline x_i\delta_{i,\min}\gamma_{i,\min}^2$, respectively. Since $\underline x_i,\delta_{i,\min},\gamma_{i,\min}\in(0,1]$, each of the four bounds above is at least $\varepsilon_i=\underline{x}_i\delta_{i,\min}^2\gamma_{i,\min}^3$.
Therefore, for $Q_i(t):=P_i(t+3)P_i(t+2)P_i(t+1)$, the $D_r$-th row satisfies $[Q_i(t)]{D_r,j}\geq\varepsilon_i$ for every $j$. Let $R_i(t):=Q_i(t)-\varepsilon_i e{D_r}\boldsymbol 1^\top$. Then $R_i(t)\geq0$ and $\|R_i(t)\|_1=1-\varepsilon_i$, so that, for every $q\in\mathbb R^{2m}$,
\be \label{eq:Q-contraction} \|Q_i(t)q\|_1\leq\varepsilon_i|\mathbf1^\top q|+(1-\varepsilon_i)\|q\|_1.\ee
This is the standard contraction mechanism associated with ergodicity coefficients of stochastic matrices \citep[Sec.~3]{ipsen2011ergodicity}.
Now define $z_i(t):=w_i(t+1)-w_i(t)$ and $v^{[k]}(t):=x^{[k]}(t+1)-x^{[k]}(t)$.
Subtracting the $w_i$ dynamics at two consecutive times gives
\beu z_i\!(t+1\!)=\! P_i\!(t+1\!)z_i(t)\!+\!(\!P_i\!(t+1\!)\!-\!P_i\!(t)\!)w_i(t)+b_i\!(t+1\!)\!-b_i\!(t). \eeu
Iterating it over three steps yields
\be \label{eq:z-three-steps} \begin{aligned}
&\mspace{-5mu}z_i (t+3)=\,Q_i(t)z_i(t) \\
&\mspace{-5mu}+P_i(t+3)P_i(t+2)\big((P_i(t+1)-P_i(t))w_i(t)+b_i(t+1)-b_i(t)\big) \\
&\mspace{-5mu}+P_i(t+3)\big(\!(\!P_i(t+2)-P_i(t+1\!)\!)w_i(t+1\!)+b_i\!(t+2)-b_i\!(t+1)\!\big) \\
&\mspace{-5mu}+\! \big(\!(P_i(t+3)-P_i(t+2)\!\!)w_i(t+2)+b_i(t+3)-b_i(t+2)\! \big), 
\vspace{-5pt}
\end{aligned}\ee
Only the $D_k$-columns of $P_i(t)$ depend on the opinions. For each such column, 
\begin{align*}
    \|P_i(t+1)-P_i(t)\|_1 &\leq2\sum\nolimits_{\ell\neq k}\gamma_i^{[\ell]}|v_i^{[\ell]}(t)| \\
    &\leq2\big(\sum\nolimits_k\!\! \gamma_i^{[k]}\big)\max_k|v_i^{[k]}(t)| \leq2\max_k|v_i^{[k]}(t)|
\end{align*}
Since $w_i(t)\geq0$ and $\|w_i(t)\|_1=1-s_i(t)\leq1$ and $\sum_k\gamma_i^{[k]}\leq1$,
 \be \label{eq:diff-p-w} \|(P_i(t+1)-P_i(t))w_i(t)\|_1\leq2 \max_k|v_i^{[k]}(t)|. \ee
Moreover, 
\be \label{eq:z-zero-mass}\mspace{-15mu} \mathbf1^\top \!\! z_i(t)=\mathbf1^\top \!\! w_i(t+1)\!-\! \mathbf1^\top \!\! w_i(t)=s_i(t)-s_i(t+1)\to0. \ee
Since $P_i(t)$ is column-stochastic, $\|P_i(t)\|_1=1$. Applying \eqref{eq:Q-contraction} to the first term in \eqref{eq:z-three-steps}, using \eqref{eq:diff-p-w}, \eqref{eq:z-zero-mass}, and $b_i(t+1)-b_i(t)\to0$, and taking $\limsup$, gives
$\limsup_{t\to\infty}\|z_i(t)\|_1 \leq(1-\varepsilon_i)\limsup_{t\to\infty}\|z_i(t)\|_1 \! +6 \max_k\limsup_{t\to\infty}\|v^{[k]}(t)\|_\infty.$
Therefore
$$\limsup_{t\to\infty}\|z_i(t)\|_1 \leq\frac{6}{\varepsilon_i}\max_k\limsup_{t\to\infty}\|v^{[k]}(t)\|_\infty.$$
Taking the $\max$ over $i$, we obtain
\be \label{eq:ZwX} \max_i \limsup_{t\to\infty}\|z_i(t)\|_1 \leq\frac{6}{\min_i \varepsilon_i}\max_k\limsup_{t\to\infty}\|v^{[k]}(t)\|_\infty.\ee
Conversely, subtracting the opinion dynamics at two consecutive times gives
$$v^{[k]}(t+1)=\Lambda^{[k]}\widetilde Wv^{[k]}(t)+\Xi^{[k]}W\big(a^{[k]}(t+1)-a^{[k]}(t)\big). $$
Since $W$ and $\widetilde W$ are row-stochastic and $\|a^{[k]}(t+1)-a^{[k]}(t)\|_\infty\leq\max_i\|z_i(t)\|_1$, taking the $\limsup$ yields
$$\limsup_{t\to\infty}\|v^{[k]}(t)\|_\infty\leq\frac{\max_i\xi_i^{[k]}}{1-\max_i\lambda_i^{[k]}} \max_i \limsup_{t\to\infty}\|z_i(t)\|_1.$$
Taking the maximum over $k$ gives
\vspace{-3pt}\be \label{eq:XwZ} \max_k\limsup_{t\to\infty}\|v^{[k]}(t)\|_\infty\leq \max_k\frac{\ds \max_i \xi_i^{[k]}}{\ds 1-\max_i \lambda_i^{[k]}} \max_i \limsup_{t\to\infty}\|z_i(t)\|_1. \vspace{-2pt}\ee
Combining \eqref{eq:ZwX} and \eqref{eq:XwZ}, condition \eqref{eq:hyp-convergence} implies
$\max_k\limsup_{t\to\infty}\|v^{[k]}(t)\|_\infty=0$,  $\limsup_{t\to\infty}\|z_i(t)\|_1=0.$
Hence, for every $i$ and $k$, $x_i^{[k]}(t+1)-x_i^{[k]}(t)\to0$, $a_i^{[k]}(t+1)-a_i^{[k]}(t)\to0$, $d_i^{[k]}(t+1)-d_i^{[k]}(t)\to0$.
Since also $s_i(t+1)-s_i(t)\to0$, we conclude that $y(t+1)-y(t)\to\boldsymbol 0$.\\
\indent It remains to characterize the limit points. Since $\X$ is compact, the trajectory admits at least one limit point $\bar y$. Since $y(t+1)-y(t)\to0$ and the dynamics are continuous, every limit point is an equilibrium.
Moreover, $s(0)\neq\boldsymbol 1$ and monotonicity of $s(t)$ exclude the AFE. Hence, by Corollary~\ref{cor:equilibrium_dichotomy}, every limit point is an ADE. Since the ADE is unique, every limit point equals $y^\star$, and therefore $y(t)\to y^\star$.

\section{Proofs of the results in Section~\ref{sec:opinion-control}}\label{apx:proof_sec_4}
\subsection{Proof of Theorem \ref{thm:existence}}\label{apx:proof_thm_existence}
The proof reduces the finite-horizon control problems to the maximization of continuous functions over compact feasible sets.
Since the horizon is finite, $\mc U$ is a finite product of the compact set $U$ and is therefore compact. The targeted-control set $\mc U_{k_0}$ is a closed subset of $\mc U$, hence compact. Moreover, the budget-feasible set
$\big\{ u\in\mc U_{k_0}: \sum\nolimits_{t=0}^{T-1}u(t)^\top Q u(t)\le B \big\}$
is a nonempty closed subset of $\mc U_{k_0}$, since the budget functional is continuous and the zero control is feasible.
For every admissible control sequence $u(\cdot)\in\mc U$, the state trajectory is uniquely determined by the $y(t+1)=f(y(t),u(t))$, with $y(0)=y_0$. Since $f$ is continuous, the finite-horizon trajectory $y(0),\ldots,y(T)$ depends continuously on $u(\cdot)$. Therefore, $J_1$ and $J_2$, being finite sums of continuous functions of the state and control variables, are continuous on their respective feasible sets. The result then follows from the Weierstrass theorem.

\bibliographystyle{agsm}
\bibliography{competing_tec}

@article{Bass1969,
	author = {Bass, Frank M.},
	title = {A New Product Growth for Model Consumer Durables},
	journal = {Manage. Sci.},
	volume = {15},
	number = {5},
	pages = {215-227},
	year = {1969}
}

@article{Pastor-Satorras2015,
	author = {Pastor-Satorras, Romualdo and Castellano, Claudio and Van Mieghem, Piet and Vespignani, Alessandro},
	title = {Epidemic processes in complex networks},
	journal = {Reviews of Modern Physics},
	volume = {87},
	number = {3},
	pages = {925-979},
	year = {2015}
}

@article{Granovetter1978,
	author = {Granovetter, Mark},
	title = {Threshold Models of Collective Behavior},
	journal = {Am. Journal of Sociology},
	volume = {83},
	number = {6},
	pages = {1420-1443},
	year = {1978}
}

@article{Friedkin1990,
	author = {Friedkin, Noah E. and Johnsen, Eugene C.},
	title = {Social Influence and Opinions},
	journal = {J. Math. Sociol.},
	volume = {15},
	number = {3-4},
	pages = {193-206},
	year = {1990}
}

@article{Kermack.McKendrick:1927,
	author = {Kermack, W. O. and McKendrick, A. G.},
	journal = {Proc. of the {R}oyal {S}ociety of {L}ondon. Series A},
	number = {772},
	pages = {700--721},
	title = {A contribution to the mathematical theory of epidemics},
	volume = {115},
	year = {1927}}

@article{Xu2024,
	author = {Q. Xu and H. Ishii},
	title = {On a Discrete-Time Networked {SIV} Epidemic Model with Polar Opinion Dynamics},
	journal = {IEEE Trans. Netw. Sci. Eng.},
	volume = {11},
	number = {6},
	pages = {6636-6651},
	year = {2024}
}

@article{DeGroot1974,
	author = {M. H. DeGroot},
	title = {Reaching a consensus},
	journal = {Journal of the Am. Stat. Ass.},
	volume = {69},
	number = {345},
	pages = {118--121},
	year = {1974}
}

@article{LI20141042,
	title = {Analysis of epidemic spreading of an {SIRS} model in complex heterogeneous networks},
	journal = {Comm. in Nonlinear Sci. and Numerical Simulation},
	volume = {19},
	number = {4},
	pages = {1042-1054},
	year = {2014},
	issn = {1007-5704},
	author = {Chun-Hsien Li and Chiung-Chiou Tsai and Suh-Yuh Yang},
}

@inproceedings{Kempe2003,
    author = {Kempe, David and Kleinberg, Jon and Tardos, \'{E}va},
    title = {Maximizing the spread of influence through a social network},
    year = {2003},
    publisher = {Ass. for Computing Machinery},
    address = {New York, NY, USA},
    booktitle = {Proc. 9th ACM SIGKDD Int. Conf. Knowl. Discov. Data Min.},
    pages = {137–146},
    numpages = {10},
}

@article{She2022,
	author    = {B. She and J. Liu and S. Sundaram and P. E. Par{\'e}},
	title     = {On a networked {SIS} epidemic model with cooperative and antagonistic opinion dynamics},
	journal   = {IEEE Trans. Control Netw. Syst.},
	volume    = {9},
	number    = {3},
	pages     = {1154--1165},
	year      = {2022},
	doi       = {},
}

@article{morris2000contagion,
	title={Contagion},
	author={Morris, Stephen},
	journal={The Review of Economic Studies},
	volume={67},
	number={1},
	pages={57--78},
	year={2000},
	publisher={Wiley-Blackwell}
}

@article{daley1964epidemics,
	title={Epidemics and rumours},
	author={Daley, Daryl J and Kendall, David G},
	journal={Nature},
	volume={204},
	number={4963},
	pages={1118--1118},
	year={1964},
	publisher={Nature Publishing Group UK London}
}

@book{Rogers2010,
	title={Diffusion of Innovations},
	author={Rogers, Everett M.},
	year={2010},
	publisher={Simon and Schuster},
	address={New York}
}

@article{bettencourt2006power,
	title={The power of a good idea: Quantitative modeling of the spread of ideas from epidemiological models},
	author={Bettencourt, Lu{\'\i}s MA and Cintr{\'o}n-Arias, Ariel and Kaiser, David I and Castillo-Ch{\'a}vez, Carlos},
	journal={Physica A: Stat. Mechanics and its Applications},
	volume={364},
	pages={513--536},
	year={2006},
	publisher={Elsevier}
}

@article{goffman1964generalization,
	title={Generalization of epidemic theory},
	author={Goffman, William and Newill, V},
	journal={Nature},
	volume={204},
	number={4955},
	pages={225--228},
	year={1964}
}

@article{li2020analysis,
	title={Analysis of the impact of policies intervention on electric vehicles adoption considering information transmission-based on consumer network model},
	author={Li, Jingjing and Jiao, Jianling and Tang, Yunshu},
	journal={Energy Policy},
	volume={144},
	pages={111560},
	year={2020},
	publisher={Elsevier}
}

@article{granell2013dynamical,
	title={Dynamical interplay between awareness and epidemic spreading in multiplex networks},
	author={Granell, Clara and G{\'o}mez, Sergio and Arenas, Alex},
	journal={Physical Review Letters},
	volume={111},
	number={12},
	pages={128701},
	year={2013},
	publisher={APS}
}

@article{wang2019impact,
	title={The impact of awareness diffusion on {SIR}-like epidemics in multiplex networks},
	author={Wang, Zhishuang and Guo, Quantong and Sun, Shiwen and Xia, Chengyi},
	journal={Applied Math. and Computation},
	volume={349},
	pages={134--147},
	year={2019},
	publisher={Elsevier}
}

@inproceedings{lin2021discrete,
	title={On a discrete-time network {SIS} model with opinion dynamics},
	author={Lin, Yixuan and Xuan, Weihao and Ren, Ruijie and Liu, Ji},
	booktitle={Proc. of the 2021 60th IEEE Conf. Decision Control},
	pages={2098--2103},
	year={2021},
}

@inproceedings{she2021peak,
	title={Peak infection time for a networked {SIR} epidemic with opinion dynamics},
	author={She, Baike and Leung, Humphrey CH and Sundaram, Shreyas and Par{\'e}, Philip E},
	booktitle={Proc. of the 2021 60th IEEE Conf. Decision Control},
	pages={2104--2109},
	year={2021},
}

@inproceedings{arango2025opinion,
	title={Opinion-driven risk perception and reaction in {SIS} epidemics},
	author={Arango, Marcela Ordorica and Bizyaeva, Anastasia and Levin, Simon A and Leonard, Naomi Ehrich},
	booktitle={Proc. of the 2025 IEEE Am. Control Conf. (ACC)},
	pages={4998--5003},
	year={2025},
}

@inproceedings{bizyaeva2024active,
	title={Active risk aversion in {SIS} epidemics on networks},
	author={Bizyaeva, Anastasia and Arango, Marcela Ordorica and Zhou, Yunxiu and Levin, Simon and Leonard, Naomi Ehrich},
	booktitle={Proc. of the IEEE 2024 Am. Control Conf. (ACC)},
	pages={4428--4433},
	year={2024},
}

@article{centola2010spread,
  title={The spread of behavior in an online social network experiment},
  author={Centola, Damon},
  journal={Science},
  volume={329},
  number={5996},
  pages={1194--1197},
  year={2010},
  publisher={Am. Ass. for the Advancement of Sci.}
}

@article{alutto2025predictive,
  title={Predictive Control Strategies for Sustaining Innovation Adoption on Multilayer Social Networks},
  author={Alutto, Martina and Xu, Qiulin and Dabbene, Fabrizio and Ishii, Hideaki and Ravazzi, Chiara},
  journal={arXiv preprint arXiv:2509.01457},
  year={2025}
}

@article{alutto2025modeling,
  title={Modeling and Control of Sustainable Transitions through Opinion-Behavior Coupling in Heterogeneous Networks},
  author={Alutto, Martina and Bellotti, Sofia and Dabbene, Fabrizio and Ravazzi, Chiara},
  journal={IEEE Transactions on Computational Social Systems},
  year={2026},
}

@inproceedings{fazeli2012game,
  title={Game theoretic analysis of a strategic model of competitive contagion and product adoption in social networks},
  author={Fazeli, Arastoo and Jadbabaie, Ali},
  booktitle={Proc. of the 2012 51st IEEE Conf. Decision Control},
  pages={74--79},
  year={2012},
}

@article{arthur1989competing,
  title={Competing technologies, increasing returns, and lock-in by historical events},
  author={Arthur, W Brian},
  journal={The Economic Journal},
  volume={99},
  number={394},
  pages={116--131},
  year={1989},
  publisher={Oxford University Press Oxford, UK}
}

@article{david1985clio,
  title={Clio and the Economics of {QWERTY}},
  author={David, Paul A},
  journal={The Am. Economic Review},
  volume={75},
  number={2},
  pages={332--337},
  year={1985},
  publisher={JSTOR}
}

@article{min2018competing,
  title={Competing contagion processes: Complex contagion triggered by simple contagion},
  author={Min, Byungjoon and San Miguel, Maxi},
  journal={Scientific Reports},
  volume={8},
  number={1},
  pages={10422},
  year={2018},
  publisher={Nature Publishing Group UK London}
}

@article{liu2019analysis, 
title={Analysis and control of a continuous-time bi-virus model}, 
author={Liu, Ji and Par{\'e}, Philip E and Nedi{\'c}, Angelia and Tang, Choon Yik and Beck, Carolyn L and Ba{\c{s}}ar, Tamer}, 
journal={IEEE Trans. Autom. Control}, 
volume={64}, 
number={12}, 
pages={4891--4906}, 
year={2019}, 
}

@article{gracy2025modeling, 
title={Modeling and analysis of a coupled {SIS} bi-virus model}, 
author={Gracy, Sebin and Par{\'e}, Philip E and Liu, Ji and Sandberg, Henrik and Beck, Carolyn L and Johansson, Karl Henrik and Ba{\c{s}}ar, Tamer}, 
journal={Automatica}, 
volume={171}, 
pages={111937}, 
year={2025}, 
publisher={Elsevier} }

@article{XUAN20202582,
title = {On a Network SIS Model with Opinion Dynamics},
journal = {IFAC-PapersOnLine},
volume = {53},
number = {2},
pages = {2582-2587},
year = {2020},
note = {21st IFAC World Congress},
issn = {2405-8963},
doi = {https://doi.org/10.1016/j.ifacol.2020.12.305},
author = {Weihao Xuan and Ruijie Ren and Philip E. Paré and Mengbin Ye and Sebastian Ruf and Ji Liu},
}

@article{de2024game,
  title={A game theory analysis of decentralized epidemic management with opinion dynamics},
  author={De Silva, Olivier Lindamulage and Lasaulce, Samson and Mor{\u{a}}rescu, Irinel-Constantin and Varma, Vineeth S},
  journal={IEEE Trans. Control Netw. Syst.},
  volume={12},
  number={1},
  pages={1137--1149},
  year={2024},
}

@article{christakis2007spread,
  title={The spread of obesity in a large social network over 32 years},
  author={Christakis, Nicholas A and Fowler, James H},
  journal={New England Journal of Medicine},
  volume={357},
  number={4},
  pages={370--379},
  year={2007},
  publisher={Mass Medical Soc}
}

@article{valente2005network,
  title={Network models and methods for studying the diffusion of innovations},
  author={Valente, Thomas W},
  journal={Models and Methods in Social Network Analysis},
  volume={28},
  pages={98--116},
  year={2005},
  publisher={Cambridge University Press New York, NY}
}

@inproceedings{ruf2019viral,
  title={A viral model of product adoption with antagonistic interactions},
  author={Ruf, Sebastian F and Par{\'e}, Philip E and Liu, Ji and Beck, Carolyn L and Ba{\c{s}}ar, Tamer},
  booktitle={Proc. IEEE 2019 Am. Control Conf. (ACC)},
  pages={3382--3387},
  year={2019},
}

@article{fagnani2017diffusion,
  title={Diffusion of innovation in large scale graphs},
  author={Fagnani, Fabio and Zino, Lorenzo},
  journal={IEEE Trans. Netw. Sci. Eng.},
  volume={4},
  number={2},
  pages={100--111},
  year={2017},
}

@article{sun2019model,
  title={A model for competing information diffusion in social networks},
  author={Sun, Qingsong and Li, Ying and Hu, Haibo and Cheng, Shulin},
  journal={IEEE Access},
  volume={7},
  pages={67916--67922},
  year={2019},
}

@article{alutto2026coupled,
  title={On a Coupled Adoption-Opinion Framework for Competing Innovations},
  author={Alutto, Martina and Dabbene, Fabrizio and Fontan, Angela and Johansson, Karl H and Ravazzi, Chiara},
  journal={Proc. of the 23rd IFAC World Congress},
  year={2026}
}

@book{bertsekas2012dynamic,
  title={Dynamic programming and optimal control: Volume I},
  author={Bertsekas, Dimitri},
  volume={4},
  year={2012},
  publisher={Athena scientific}
}

@article{mcasey2012convergence,
  title={Convergence of the forward-backward sweep method in optimal control},
  author={McAsey, Michael and Mou, Libin and Han, Weimin},
  journal={Comput. Optim. Appl.},
  volume={53},
  number={1},
  pages={207--226},
  year={2012},
  publisher={Springer}
}

@article{ipsen2011ergodicity,
  title={Ergodicity coefficients defined by vector norms},
  author={Ipsen, Ilse CF and Selee, Teresa M},
  journal={SIAM J. Matrix Anal. Appl.},
  volume={32},
  number={1},
  pages={153--200},
  year={2011},
}

@article{katona2011network,
  author  = {Katona, Zsolt and Zubcsek, Peter Pal and Sarvary, Miklos},
  title   = {Network Effects and Personal Influences:
             The Diffusion of an Online Social Network},
  journal = {Journal of Marketing Research},
  volume  = {48},
  number  = {3},
  pages   = {425--443},
  year    = {2011},
  doi     = {10.1509/jmkr.48.3.425}
}

@article{katz1986technology,
  author  = {Katz, Michael L. and Shapiro, Carl},
  title   = {Technology Adoption in the Presence of Network Externalities},
  journal = {Journal of Polit. Econ.},
  volume  = {94},
  number  = {4},
  pages   = {822--841},
  year    = {1986},
  doi     = {10.1086/261409}
}

@article{katz1992product,
  author  = {Katz, Michael L. and Shapiro, Carl},
  title   = {Product Introduction with Network Externalities},
  journal = {The Journal of Industrial Economics},
  volume  = {40},
  number  = {1},
  pages   = {55--83},
  year    = {1992},
  doi     = {10.2307/2950627}
}

@article{kenney2011structuring,
  author  = {Kenney, Martin and Pon, Bryan},
  title   = {Structuring the Smartphone Industry:
             Is the Mobile Internet OS Platform the Key?},
  journal = {Journal of Industry, Competition and Trade},
  volume  = {11},
  pages   = {239--261},
  year    = {2011},
  doi     = {10.1007/s10842-011-0105-6}
}

@article{aral2011creating,
  author  = {Aral, Sinan and Walker, Dylan},
  title   = {Creating Social Contagion Through Viral Product Design:
             A Randomized Trial of Peer Influence in Networks},
  journal = {Manage. Sci.},
  volume  = {57},
  number  = {9},
  pages   = {1623--1639},
  year    = {2011},
  doi     = {10.1287/mnsc.1110.1421}
}

\end{document}